\documentclass[11pt,a4paper]{article}
\usepackage{jheppub}
\usepackage{bbm}
\usepackage{bm}
\usepackage{cancel}
\usepackage{mathrsfs}

\let\de=\partial
\newcommand\dd{\mathrm{d}}
\newcommand\g{\mathfrak{g}}
\newcommand\h{\mathfrak{h}}
\newcommand\e{\mathfrak{e}}
\newcommand\uu{\mathfrak{u}}
\newcommand\imag{\mathrm{i}}
\newcommand\openone{\mathbbm{1}}
\newcommand\La{\mathscr{L}}
\newcommand\pCPT{\Phi}
\newcommand\BCPT{\mathcal{G}}
\newcommand\CPT{\mathcal U}
\newcommand\gr[1]{\mathrm{#1}}
\newcommand\gtr[2]{\mathcal T_{#1}{#2}}
\newcommand\cst[1]{\chi_{#1}}
\newcommand\cpr[2]{(#1,#2)}
\DeclareMathOperator\tr{Tr}
\newtheorem{theorem}{Theorem}
\definecolor{myred}{rgb}{1,0,0}
\definecolor{mygreen}{rgb}{0,0.5,0.2}
\definecolor{myblue}{rgb}{0,0,1}
\newcommand{\oned}[1]{{\color{mygreen}#1}}
\newcommand{\twod}[1]{{\color{myred}#1}}
\newcommand{\threed}[1]{{\color{myblue}#1}}

\title{Effective Lagrangians for quantum many-body systems}

\author[a]{Jens O.~Andersen,}
\author[b,c]{Tom\'{a}\v{s} Brauner,}
\author[d]{Christoph P.~Hofmann}
\author[e]{and Aleksi Vuorinen}

\affiliation[a]{Department of Physics, Norwegian University of Science and Technology, Trondheim, Norway}
\affiliation[b]{Institute for Theoretical Physics, Vienna University of Technology, Vienna, Austria}
\affiliation[c]{Department of Theoretical Physics, Nuclear Physics Institute of  the ASCR, \v{R}e\v{z}, Czech Republic}
\affiliation[d]{Facultad de Ciencias, Universidad de Colima, Colima, Mexico}
\affiliation[e]{Department of Physics and Helsinki Institute of Physics, University of Helsinki, Helsinki, Finland}

\emailAdd{jens.andersen@ntnu.no}
\emailAdd{brauner@hep.itp.tuwien.ac.at}
\emailAdd{christoph.peter.hofmann@gmail.com}
\emailAdd{aleksi.vuorinen@helsinki.fi}

\abstract{The low-energy and low-momentum dynamics of systems with a spontaneously broken continuous symmetry is dominated by the ensuing Nambu-Goldstone bosons. It can be conveniently encoded in a model-independent effective field theory whose structure is fixed by symmetry up to a set of effective coupling constants. We construct the most general effective Lagrangian for the Nambu-Goldstone bosons of spontaneously broken global internal symmetry up to the fourth order in derivatives. Rotational invariance and spatial dimensionality of one, two or three are assumed in order to obtain compact explicit expressions, but our method is completely general and can be applied without modifications to condensed matter systems with a discrete space group as well as to higher-dimensional theories. The general low-energy effective Lagrangian for relativistic systems follows as a special case. We also discuss the effects of explicit symmetry breaking and classify the corresponding terms in the Lagrangian. Diverse examples are worked out in order to make the results accessible to a wide theoretical physics community.}


\begin{document}
\maketitle
\flushbottom


\section{Introduction}
\label{sec:intro}

The methods of effective field theory (EFT) have proven invaluable across a range of disciplines as a tool for simplifying practical calculations in systems with two or more widely separated energy or length scales~\cite{Georgi:1994qn,Manohar:1996cq}. Physical observables at long distances can be determined using an EFT that respects the correct symmetries  and only includes the low-energy degrees of freedom. The effects of microscopic, short-distance interactions are then encoded in a set of effective coupling constants that are determined by experiment or computed from an underlying fundamental theory.

A major problem in the construction of an EFT is the choice of the appropriate degrees of freedom. Fortunately, there is a large class of physical systems where this task can be accomplished at once: whenever a continuous symmetry is spontaneously broken, the spectrum of the theory contains gapless excitations, the Nambu-Goldstone (NG) bosons. Examples of NG bosons include sound waves --- the phonons --- in solids and (super)fluids, spin waves --- the magnons --- in (anti)ferromagnets, or pions in quantum chromodynamics (QCD). Provided there are no other soft modes in the spectrum, not associated with symmetry, the low-energy dynamics is dominated by the NG bosons alone. This assumption will be implicit throughout the remainder of the paper.

The formalism of EFT for the NG bosons of a spontaneously broken symmetry was developed in full generality in high energy physics. In particular, Coleman et al.~\cite{Coleman:1969sm,Callan:1969sn} showed how to construct effective Lagrangians for the NG bosons, invariant under an arbitrary compact internal symmetry group. As a consequence of the spontaneously broken symmetry, the NG bosons interact weakly at low energy or momentum, and the EFT Lagrangian can be organized as a series of terms with an increasing number of derivatives~\cite{Weinberg:1978kz}. A prime example of the application of EFT methods to a precision analysis of low-energy dynamics is chiral perturbation theory ($\chi$PT) of QCD~\cite{Gasser:1983yg,Gasser:1984gg}. For nonrelativistic, condensed matter systems, the use of EFT techniques was on the other hand advocated by Leutwyler~\cite{Leutwyler:1993gf}, who developed a general EFT framework to leading, second order in derivatives. Detailed applications including selected higher-order calculations were subsequently worked out for the special cases of ferromagnets~\cite{Roman:1999ro,Roman:canted,Hofmann:2011dm,Hofmann:2012ms,Hofmann:2013fqa} and antiferromagnets~\cite{Hofmann:1997qm,Kampfer:2005ba,Brugger:2006dz,Hofmann:2009ru}.

Despite these examples, the application of EFT to NG bosons in nonrelativistic systems has not been developed to the same extent as in relativistic field theory. In Lorentz-invariant systems in four spacetime dimensions and provided quantum anomalies are absent, the effective Lagrangian can be assumed to be invariant with respect to the relevant symmetries without loss of generality~\cite{Leutwyler:1993iq}, and the methods of refs.~\cite{Coleman:1969sm,Callan:1969sn} can thus be used to construct it. Once Lorentz invariance is given up, the Lagrangian, however, becomes in general invariant only up to a total derivative. While this still guarantees the invariance of the action, it makes its explicit construction a nontrivial problem. In his seminal paper, Leutwyler~\cite{Leutwyler:1993gf} derived a set of differential equations for the nonlinear dependence of the leading-order effective Lagrangian on the NG fields, dictated by symmetry. In addition, he found their solution for the special case of an (anti)ferromagnet.

A general solution to Leutwyler's differential equations was discovered only recently~\cite{Watanabe:2013uya} (see also ref.~\cite{Watanabe:2014fva} for a more detailed discussion). The objective of the present paper is to fill a gap in the development of EFT and show, assuming absence of anomalies, how the construction of the effective Lagrangian can be carried out at higher orders in the derivative expansion. There are good reasons to be concerned with higher-order contributions, despite the computational complexity that accompanies such an analysis. The first one is precision, of which the calculation of selected observables in $\chi$PT to the sixth order in derivatives sets an example~\cite{Bijnens:2006zp}. Perhaps more importantly, the leading-order Lagrangian often possesses an accidental symmetry which is not inherent to the microscopic theory. Higher-order operators can then actually provide a dominant contribution to certain rare processes~\cite{Witten:1983tw}. Finally, higher-order operators are needed as counterterms whenever loops are taken into account, which is a necessity if one wishes to discuss the thermodynamics of broken symmetry~\cite{Gasser:1987ah,Splittorff:2001fy,Splittorff:2002xn}.

The main results of this paper are: (i) classification of all terms in the effective Lagrangian, to order four in the gradient expansion, that are invariant up to a total derivative, announced in ref.~\cite{CSpaper}; (ii) a transparent algorithm for the construction of all the remaining, strictly invariant terms in the Lagrangian. We moreover provide an explicit expression for the most general effective Lagrangian up to order four in derivatives, assuming for simplicity rotational invariance. The effects of explicit symmetry breaking are also discussed.


\subsection{Plan of the paper}
\label{subsec:plan}

Our ultimate aim is to provide a systematic framework suitable for applications in both high energy and condensed matter physics. This determines the structure of the paper. In section~\ref{sec:summary}, we summarize our results, introducing only the minimum amount of notation necessary. In order to make the complicated expressions more intelligible, we furthermore highlight contributions to the effective Lagrangian that are specific to certain spatial dimensions or that encode explicit symmetry breaking. Some concrete examples are subsequently worked out in section~\ref{sec:examples} to further clarify the formalism. These two sections constitute the essence of the paper, necessary for the reader interested in practical applications rather than general developments.

After introducing the practical results, the rest of the paper presents the conceptual background behind the construction. Following largely the foundational work of Leutwyler~\cite{Leutwyler:1993iq}, section~\ref{sec:method} explains how the construction of the effective Lagrangian can be reduced to an elementary problem in field theory. As a warmup and for illustration, we show in section~\ref{sec:LO} how the leading-order Lagrangian of ref.~\cite{Leutwyler:1993gf}, including the explicit solution for its coupling functions~\cite{Watanabe:2013uya}, is reproduced elegantly in our approach. We also derive the corresponding equation of motion, which can be used to eliminate some of the operators at higher orders. Section~\ref{sec:higherorders} then provides some details of the construction of effective Lagrangians at the next two orders of the derivative expansion. In particular, section~\ref{subsec:invariant} deals with the invariant part of the Lagrangian. The most subtle part of the construction, namely the classification of terms invariant only up to a total derivative~\cite{CSpaper}, is reviewed in detail in sections~\ref{subsec:chernsimons} and~\ref{subsec:CSphysical}. Finally, the effects of explicit symmetry breaking are discussed in section~\ref{subsec:explicitSB}. Although most of the technical details are provided in the main body of the paper, some auxiliary results that can be formulated separately are deferred to the appendices.


\section{Summary of the results}
\label{sec:summary}

\subsection{Setup and notation}
\label{subsec:setup}

To facilitate the unique definition of the effective Lagrangian, we first introduce the most important notation necessary. More detailed properties of the individual building blocks will be discussed below.
\begin{itemize}
\itemsep=-2pt
\item Internal symmetry group: $G$.
\item Corresponding symmetry generators: $T_{i,j,k,\dotsc}$.
\item Unbroken subgroup: $H$.
\item Unbroken generators: $T_{\alpha,\beta,\gamma,\dotsc}$.
\item Broken generators: $T_{a,b,c,\dotsc}$.
\item Structure constants: $f^k_{ij}$; defined by $[T_i,T_j]=\imag f^k_{ij}T_k$; $f^a_{\alpha\beta}$ always vanishes; $f^\alpha_{\beta a}=0$ is assumed (can be ensured by a suitable choice of basis for all compact Lie algebras).
\item Spacetime indices: $\kappa,\lambda,\mu,\nu,\dotsc$.
\item Spatial indices: $r,s,t,\dotsc$.
\item Nambu-Goldstone fields: $\pi^a$; encoded in a matrix variable $U(\pi)$; parameterization  arbitrary except for the requirement that the vacuum corresponds to $\pi=0$, $U(0)=\openone$.
\item External gauge fields: $A^i_\mu$.
\item External gauge field-strength tensor: $F^i_{\mu\nu}=\de_\mu A^i_\nu-\de_\nu A^i_\mu+f^i_{jk}A^j_\mu A^k_\nu$.
\item Auxiliary field variables: $\phi^a_\mu$, $B^\alpha_\mu$; defined by $U^{-1}(A^i_\mu T_i)U+\imag U^{-1}\de_\mu U=\phi^a_\mu T_a+B^\alpha_\mu T_\alpha$.
\item Auxiliary field covariant derivative: $D_\mu\phi_\nu^a=\de_\mu\phi_\nu^a+f^a_{\alpha b}B^\alpha_\mu\phi^b_\nu$.
\item Auxiliary field-strength tensor: $G^\alpha_{\mu\nu}=\de_\mu B^\alpha_\nu-\de_\nu B^\alpha_\mu+f^\alpha_{\beta\gamma}B^\beta_\mu B^\gamma_\nu$.
\item Explicit symmetry breaking parameter: $m_{\rho,\sigma,\dotsb}$; enters the microscopic theory through the operator $m_\sigma\mathcal O^\sigma$.
\item Auxiliary mass field: $\Xi_\sigma$; defined by $\Xi_\sigma=D(U)^\rho_{\phantom\rho\sigma}m_\rho$, where $D$ is the representation of the symmetry group in which $\mathcal O^\sigma$ transforms.
\end{itemize}


\subsection{Effective Lagrangian}
\label{subsec:lagrangian}

To the order that we are interested in, the effective Lagrangian takes the form of a polynomial in the auxiliary fields $\phi^a_\mu$ and $B^\alpha_\mu$, the field-strength tensor $G^\alpha_{\mu\nu}$ and the covariant derivative $D_\mu\phi^a_\nu$. It is written as a sum $\La_\text{eff}=\La_\text{inv}+\La_\text{CS}+\La_\text{s.b.}$. The first part here is strictly invariant under simultaneous gauge transformations of the NG and external gauge fields, while the second part is only invariant up to a surface term. Finally, the third part incorporates the effects of explicit symmetry breaking. Each part of the Lagrangian can be further organized as a sum of contributions $\La^{(s,t)}$, carrying $s$ spatial and $t$ temporal indices. In practice, this splitting is only necessary for $\La_\text{inv}$ which includes a large number of terms.

For the sake of simplicity, we assume invariance under \emph{continuous spatial rotations}. The same approach can, however, be applied without modifications to arbitrary spacetime symmetry. Fully general expressions for $\La_\text{inv}^{(s,t)}$ with $s+t\leq2$ and for $\La_\text{CS}^{(s,t)}$ with $s+t\leq4$ are given in sections~\ref{sec:LO} and~\ref{subsec:chernsimons}, respectively.


\subsubsection{Invariant part of the Lagrangian}
\label{subsubsec:Linvar}

Here, we list all operators that appear in $\La_\text{inv}^{(s,t)}$ with $s+t\leq4$, modulo ambiguities due to integration by parts. As some operators exist in any spacetime dimensionality while others do not, we use color coding to highlight operators particular to \oned{one}, \twod{two} and \threed{three} \emph{spatial} dimensions. Also, we list separately operators containing the field-strength tensor $G^\alpha_{\mu\nu}$.
\begin{itemize}
\item[$\La_\text{inv}^{(0,1)}$:] $\phi^a_0$.

\item[$\La_\text{inv}^{(1,0)}$:] $\oned{\phi^a_1}$.

\item[$\La_\text{inv}^{(0,2)}$:] $\phi^a_0\phi^b_0$.

\item[$\La_\text{inv}^{(1,1)}$:] $\oned{\phi^a_0\phi^b_1}$.

\item[$\La_\text{inv}^{(2,0)}$:] $\phi^a_r\phi^b_r$, $\twod{\epsilon^{rs}\phi^a_r\phi^b_s}$.

\item[$\La_\text{inv}^{(0,3)}$:] $\phi^a_0\phi^b_0\phi^c_0$, $\phi^a_0D_0\phi^b_0$.

\item[$\La_\text{inv}^{(1,2)}$:] $\oned{\phi^a_0\phi^b_0\phi^c_1}$, $\oned{\phi^a_0D_0\phi^b_1}$, $\oned{\phi^a_0D_1\phi^b_0}$,\\[1ex]
$\oned{\phi^a_0G^\alpha_{01}}$.

\item[$\La_\text{inv}^{(2,1)}$:] $\phi^a_0\phi^b_r\phi^c_r$, $\twod{\epsilon^{rs}\phi^a_0\phi^b_r\phi^c_s}$, $\phi^a_0D_r\phi^b_r$, $\phi^a_rD_0\phi^b_r$, $\twod{\epsilon^{rs}\phi^a_0D_r\phi^b_s}$, $\twod{\epsilon^{rs}\phi^a_rD_0\phi^b_s}$,\\[1ex]
$\phi^a_rG^\alpha_{0r}$, $\twod{\epsilon^{rs}\phi^a_0G^\alpha_{rs}}$, $\twod{\epsilon^{rs}\phi^a_rG^\alpha_{0s}}$.

\item[$\La_\text{inv}^{(3,0)}$:] $\oned{\phi^a_1\phi^b_1\phi^c_1}$, $\threed{\epsilon^{rst}\phi^a_r\phi^b_s\phi^c_t}$, $\oned{\phi^a_1D_1\phi^b_1}$, $\threed{\epsilon^{rst}\phi^a_rD_s\phi^b_t}$,\\[1ex]
$\threed{\epsilon^{rst}\phi^a_rG^\alpha_{st}}$.

\item[$\La_\text{inv}^{(0,4)}$:] $\phi^a_0\phi^b_0\phi^c_0\phi^d_0$, $\phi^a_0\phi^b_0D_0\phi^c_0$, $D_0\phi^a_0D_0\phi^b_0$.

\item[$\La_\text{inv}^{(1,3)}$:] $\oned{\phi^a_0\phi^b_0\phi^c_0\phi^d_1}$, $\oned{\phi^a_0\phi^b_0D_1\phi^c_0}$, $\oned{\phi^a_0\phi^b_1D_0\phi^c_0}$, $\oned{D_0\phi^a_0D_0\phi^b_1}$, $\oned{D_0\phi^a_0D_1\phi^b_0}$,\\[1ex]
$\oned{\phi^a_0\phi^b_0G^\alpha_{01}}$, $\oned{D_0\phi^a_0G^\alpha_{01}}$.

\item[$\La_\text{inv}^{(2,2)}$:] $\phi^a_0\phi^b_0\phi^c_r\phi^d_r$, $\twod{\epsilon^{rs}\phi^a_0\phi^b_0\phi^c_r\phi^d_s}$, $\phi^a_0\phi^b_rD_0\phi^c_r$, $\twod{\epsilon^{rs}\phi^a_0\phi^b_rD_0\phi^c_s}$, $\phi^a_r\phi^b_0D_r\phi^c_0$, $\twod{\epsilon^{rs}\phi^a_r\phi^b_0D_s\phi^c_0}$, $D_0\phi^a_rD_0\phi^b_r$, $\twod{\epsilon^{rs}D_0\phi^a_rD_0\phi^b_s}$, $D_r\phi^a_0D_r\phi^b_0$, $D_0\phi^a_0D_r\phi^b_r$, $\twod{\epsilon^{rs}D_0\phi^a_0D_r\phi^b_s}$,\\[1ex]
$\twod{\epsilon^{rs}\phi^a_0\phi^b_0G^\alpha_{rs}}$, $\phi^a_0\phi^b_rG^\alpha_{0r}$, $\twod{\epsilon^{rs}\phi^a_0\phi^b_rG^\alpha_{0s}}$, $\twod{\epsilon^{rs}D_0\phi^a_0G^\alpha_{rs}}$, $D_0\phi^a_rG^\alpha_{0r}$, $\twod{\epsilon^{rs}D_0\phi^a_rG^\alpha_{0s}}$, $D_r\phi^a_0G^\alpha_{0r}$, $G^\alpha_{0r}G^\beta_{0r}$, $\twod{\epsilon^{rs}G^\alpha_{0r}G^\beta_{0s}}$.

\item[$\La_\text{inv}^{(3,1)}$:] $\oned{\phi^a_0\phi^b_1\phi^c_1\phi^d_1}$, $\threed{\epsilon^{rst}\phi^a_0\phi^b_r\phi^c_s\phi^d_t}$, $\oned{\phi^a_0\phi^b_1D_1\phi^c_1}$, $\threed{\epsilon^{rst}\phi^a_0\phi^b_rD_s\phi^c_t}$, $\oned{\phi^a_1\phi^b_1D_0\phi^c_1}$, $\threed{\epsilon^{rst}\phi^a_r\phi^b_sD_0\phi^c_t}$, $\oned{D_1\phi^a_0D_1\phi^b_1}$, $\oned{D_0\phi^a_1D_1\phi^b_1}$, $\threed{\epsilon^{rst}D_0\phi^a_rD_s\phi^b_t}$,\\[1ex]
$\threed{\epsilon^{rst}\phi^a_0\phi^b_rG^\alpha_{st}}$, $\oned{\phi^a_1\phi^b_1G^\alpha_{01}}$, $\threed{\epsilon^{rst}\phi^a_r\phi^b_sG^\alpha_{0t}}$, $\threed{\epsilon^{rst}D_0\phi^a_rG^\alpha_{st}}$, $\oned{D_1\phi^a_1G^\alpha_{01}}$.

\item[$\La_\text{inv}^{(4,0)}$:] $\phi^a_r\phi^b_r\phi^c_s\phi^d_s$, $\twod{\epsilon^{st}\phi^a_r\phi^b_r\phi^c_s\phi^d_t}$, $\phi^a_r\phi^b_sD_r\phi^c_s$, $\twod{\epsilon^{st}\phi^a_r\phi^b_sD_r\phi^c_t}$, $\twod{\epsilon^{st}\phi^a_s\phi^b_rD_t\phi^c_r}$,$D_r\phi^a_sD_r\phi^b_s$, $\twod{\epsilon^{st}D_r\phi^a_sD_r\phi^b_t}$, $D_r\phi^a_rD_s\phi^b_s$, $\twod{\epsilon^{st}D_r\phi^a_rD_s\phi^b_t}$,\\[1ex]
$\twod{\epsilon^{st}\phi^a_r\phi^b_rG^\alpha_{st}}$, $\phi^a_r\phi^b_sG^\alpha_{rs}$, $\twod{\epsilon^{st}D_r\phi^a_rG^\alpha_{st}}$, $D_r\phi^a_sG^\alpha_{rs}$, $G^\alpha_{rs}G^\beta_{rs}$.
\end{itemize}
Each of the operators listed above comes with an effective coupling that contracts all the internal group indices carried by the operator, as in $c_{ab\alpha}\phi^a_0\phi^b_rG^\alpha_{0r}$. Each of the couplings $c_{ab\dotsb,\alpha\beta\dotsb}$ is required to be an invariant tensor of the unbroken subgroup $H$; for all allowed values of the indices, it therefore has to satisfy the constraint
\begin{equation}
c_{cb\dotsb,\alpha\beta\dotsb}f^c_{\gamma a}+c_{ac\dotsb,\alpha\beta\dotsb}f^c_{\gamma b}+\dotsb+c_{ab\dotsb,\delta\beta\dotsb}f^\delta_{\gamma\alpha}+c_{ab\dotsb,\alpha\delta\dotsb}f^\delta_{\gamma\beta}+\dotsb=0.
\label{invariancecondition}
\end{equation}
We do not attempt to find a general solution to these constraints, but leave them to be addressed case by case using tensor methods~\cite{Cvitanovic}. The simplest examples of couplings with one and two indices that occur repeatedly throughout this paper are discussed to some extent in appendix~\ref{app:tensor}. 

The lowest-order Lagrangians, with $s+t\leq2$, are well-known by now. The special case of rotationally invariant theories in three spatial dimensions was addressed already in ref.~\cite{Leutwyler:1993gf}; the full nonlinear dependence of the associated Lagrangian on the NG fields was found recently in ref.~\cite{Watanabe:2013uya}. The cases of one and two spatial dimensions are discussed in ref.~\cite{Watanabe:2014fva}. The fully general lowest-order Lagrangian, obtained with no assumptions on the spacetime symmetry, is given below in section~\ref{sec:LO}, where we also discuss its physical implications in more detail. Specific examples of higher-order Lagrangians, including the corresponding invariant couplings, are finally worked out in section~\ref{sec:examples}.


\subsubsection{Chern-Simons terms}
\label{subsubsec:LCS}

The contributions to the Lagrangian invariant up to a surface term are most easily organized by the total order in derivatives, $s+t$. It turns out that up to order four, only two types of such terms exist, one at the first and another at the third order, 
\begin{equation}
\begin{split}
\La_\text{CS}^{(1)}&=e_\alpha B^\alpha_0,\\
\La_\text{CS}^{(3)}&=c_{\alpha\beta}\epsilon^{\lambda\mu\nu}B^\alpha_\lambda(\de_\mu B^\beta_\nu+\tfrac13f^\beta_{\gamma\delta}B^\gamma_\mu B^\delta_\nu),\qquad\text{where}\quad c_{\alpha\beta}=c_{\beta\alpha}.
\end{split}
\label{CSterms}
\end{equation}
While $\La^{(1)}_\text{CS}$ exists regardless of the spacetime dimension, $\La^{(3)}_\text{CS}$ is only allowed in two or three spatial dimensions. In the latter case, the indices $\lambda,\mu,\nu$ should be interpreted as purely spatial ones. The effective couplings $e_\alpha$ and $c_{\alpha\beta}$ are again invariant tensors of the unbroken subgroup $H$, but this time with a straightforward interpretation. First, there is one free parameter $e_\alpha$ for every $\gr{U(1)}$ factor of $H$, corresponding to the vacuum expectation value of the associated conserved charge density. Second, $c_{\alpha\beta}$ is proportional to the Killing form on every simple factor of $H$, and thus contains one free parameter for each such factor. The Chern-Simons terms have distinct topological properties, in which they substantially differ from the invariant part of the effective Lagrangian, and moreover they give rise to specific interactions amongst the NG bosons. Both of these features are discussed in detail in the companion paper~\cite{CSpaper}.


\subsubsection{Effects of explicit symmetry breaking}
\label{subsubsec:Lsb}

Precisely which explicit-symmetry-breaking operators appear at a given order of the derivative expansion depends on the order that one assigns to the parameters $m_\sigma$ in the Lagrangian. We adhere to the usual practice and count $m_\sigma$ as a quantity of order two in derivatives, which follows from the fact that the kinetic term of the NG bosons typically acquires a contribution linear in $m_\sigma$. Hence, determining the action to order four requires classifying all terms in the Lagrangian with $s+t\leq2$: 
\begin{itemize}
\item[$\La_\text{s.b.}^{(0,0)}$:] $\Xi_\sigma$, $\Xi_\rho\Xi_\sigma$.

\item[$\La_\text{s.b.}^{(0,1)}$:] $\Xi_\sigma\phi^a_0$.

\item[$\La_\text{s.b.}^{(1,0)}$:] $\oned{\Xi_\sigma\phi^a_1}$.

\item[$\La_\text{s.b.}^{(0,2)}$:] $\Xi_\sigma\phi^a_0\phi^b_0$, $\Xi_\sigma D_0\phi^a_0$.

\item[$\La_\text{s.b.}^{(1,1)}$:] $\oned{\Xi_\sigma\phi^a_0\phi^b_1}$, $\oned{\Xi_\sigma D_0\phi^a_1}$, $\oned{\Xi_\sigma D_1\phi^a_0}$, $\oned{\Xi_\sigma G^\alpha_{01}}$.

\item[$\La_\text{s.b.}^{(2,0)}$:] $\Xi_\sigma\phi^a_r\phi^b_r$, $\twod{\epsilon^{rs}\Xi_\sigma\phi^a_r\phi^b_s}$, $\Xi_\sigma D_r\phi^a_r$, $\twod{\epsilon^{rs}\Xi_\sigma D_r\phi^a_s}$, $\twod{\epsilon^{rs}\Xi_\sigma G^\alpha_{rs}}$.
\end{itemize}
These operators again come with effective couplings that now include one or two indices of the type $\sigma$. The couplings $c^{\rho\sigma\dotsb}_{ab\dotsb,\alpha\beta\dotsb}$ are invariant tensors of $H$ and satisfy a relation
\begin{equation}
\begin{split}
&c^{\rho\sigma\dotsb}_{cb\dotsb,\alpha\beta\dotsb}f^c_{\gamma a}+c^{\rho\sigma\dotsb}_{ac\dotsb,\alpha\beta\dotsb}f^c_{\gamma b}+\dotsb+c^{\rho\sigma\dotsb}_{ab\dotsb,\delta\beta\dotsb}f^\delta_{\gamma\alpha}+c^{\rho\sigma\dotsb}_{ab\dotsb,\alpha\delta\dotsb}f^\delta_{\gamma\beta}+\dotsb\\
&+\imag c^{\omega\sigma\dotsb}_{ab\dotsb,\alpha\beta\dotsb}D(T_\gamma)^\rho_{\phantom\rho\omega}+\imag c^{\rho\omega\dotsb}_{ab\dotsb,\alpha\beta\dotsb}D(T_\gamma)^\sigma_{\phantom\sigma\omega}+\dotsb=0,
\end{split}
\label{invariancecondgeneral}
\end{equation}
generalizing the earlier eq.~\eqref{invariancecondition}. Concrete examples of the above operators and couplings will be discussed in section~\ref{sec:examples}.


\subsubsection{Lorentz-invariant Lagrangians}
\label{subsubsec:LLI}

Above, we have listed all terms in the effective Lagrangian allowed by rotational invariance. Relativistic Lagrangians, invariant under the full Lorentz group, are in principle a special case thereof. However, since we treated spatial and temporal indices separately, Lorentz invariance will only be reflected implicitly, in a set of linear constraints on the effective couplings. For the reader's convenience, we will now explicitly spell out the resulting Lagrangian using the usual Lorentz-covariant notation. As space and time are mixed by Lorentz transformations, the individual contributions are organized by the total degree in derivatives, $s+t$. This time, we only consider the special cases of two and three spatial dimensions, since in one-dimensional Lorentz-invariant systems spontaneous symmetry breaking is prohibited by the Coleman theorem~\cite{Coleman:1973ci}. The result reads:
\begin{itemize}
\item[$\La_\text{inv}^{(2)}$:] $\phi^a_\mu\phi^{b\mu}$.

\item[$\La_\text{inv}^{(3)}$:] $\twod{\epsilon^{\lambda\mu\nu}\phi^a_\lambda\phi^b_\mu\phi^c_\nu}$, $\twod{\epsilon^{\lambda\mu\nu}\phi^a_\lambda D_\mu\phi^b_\nu}$, $\twod{\epsilon^{\lambda\mu\nu}\phi^a_\lambda G^\alpha_{\mu\nu}}$.

\item[$\La_\text{inv}^{(4)}$:] $\phi^a_\mu\phi^{b\mu}\phi^c_\nu\phi^{d\nu}$, $\threed{\epsilon^{\kappa\lambda\mu\nu}\phi^a_\kappa\phi^b_\lambda\phi^c_\mu\phi^d_\nu}$, $\phi^{a\mu}\phi^{b\nu}D_\mu\phi^c_\nu$, $\threed{\epsilon^{\kappa\lambda\mu\nu}\phi^a_\kappa\phi^b_\lambda D_\mu\phi^c_\nu}$, $D_\mu\phi^a_\nu D^\mu\phi^{b\nu}$, $D_\mu\phi^{a\mu}D_\nu\phi^{b\nu}$,\\[1ex]
$\phi^{a\mu}\phi^{b\nu}G^\alpha_{\mu\nu}$, $\threed{\epsilon^{\kappa\lambda\mu\nu}\phi^a_\kappa\phi^b_\lambda G^\alpha_{\mu\nu}}$, $D^\mu\phi^{a\nu}G^\alpha_{\mu\nu}$, $G^\alpha_{\mu\nu}G^{\beta\mu\nu}$.

\item[$\La_\text{CS}$:] $\twod{\epsilon^{\lambda\mu\nu}B^\alpha_\lambda(\de_\mu B^\beta_\nu+\tfrac13f^\beta_{\gamma\delta}B^\gamma_\mu B^\delta_\nu)}$.

\item[$\La_\text{s.b.}$:] $\Xi_\sigma$, $\Xi_\rho\Xi_\sigma$, $\Xi_\sigma\phi^a_\mu\phi^{b\mu}$, $\Xi_\sigma D_\mu\phi^{a\mu}$.
\end{itemize}
The associated effective couplings have to satisfy the same invariance conditions as before, see~eqs.~\eqref{invariancecondition} and~\eqref{invariancecondgeneral}. The presence of a single term in the Chern-Simons sector indicates that, as shown in ref.~\cite{Leutwyler:1993iq}, in three spatial dimensions the effective Lagrangian can be made strictly gauge-invariant by a proper choice of field variables and transformation rules.


\subsection{Expansion in Nambu-Goldstone fields}
\label{subsec:piexpansion}

The effective Lagrangians listed above are expressed exclusively in terms of the auxiliary fields $\phi^a_\mu$ and $B^\alpha_\mu$. This is both an advantage and a drawback. On the one hand, we are able to write the allowed interaction terms in a very compact way, largely independent of the chosen parameterization for the NG fields. On the other hand, the implications for the actual dynamics of the NG bosons may be somewhat obscured by this economic notation. We wish to ameliorate the latter deficiency by providing here some explicit expressions in terms of the NG fields $\pi^a$. To this end, we first introduce the Maurer-Cartan (MC) form $\omega^i_a(\pi)$ and the rotation matrix $\nu^i_j(\pi)$, defined by
\begin{equation}
\omega_a(\pi)=\omega^i_a(\pi)T_i=-\imag U(\pi)^{-1}\de_a U(\pi),\qquad\nu_j(\pi)=\nu^i_j(\pi)T_i=U(\pi)^{-1}T_jU(\pi),
\label{MCform}
\end{equation}
where $\de_a=\de/\de\pi^a$. In terms of these objects, our auxiliary fields read by construction
\begin{equation}
\phi^a_\mu(\pi)=A^i_\mu\nu^a_i(\pi)-\omega^a_b(\pi)\de_\mu\pi^b,\qquad B^\alpha_\mu(\pi)=A^i_\mu\nu^\alpha_i(\pi)-\omega^\alpha_a(\pi)\de_\mu\pi^a,
\label{auxfields}
\end{equation}
which can be viewed as an expanded form of the simple matrix relation
\begin{equation}
U^{-1}A_\mu U+\imag U^{-1}\partial_\mu U=\phi_\mu+B_\mu=\phi^a_\mu T_a+B^\alpha_\mu T_\alpha,
\label{Bphidefprel}
\end{equation}
 where $A_\mu=A^i_\mu T_i$; see section~\ref{subsec:generalactions} below for a justification of this definition. Let us now choose a specific, widely used parameterization for the NG field matrix, $U(\pi)=e^{\imag\pi^aT_a}$. The virtue of the exponential parameterization is that both the MC form and the rotation matrix $\nu^i_j(\pi)$ can be easily evaluated in power series expansions up to any desired order in the NG fields,
\begin{equation}
\begin{split}
\omega^i_a(\pi)&=\delta^i_a-\frac12f^i_{ab}\pi^b+\frac16f^j_{ab}f^i_{jc}\pi^b\pi^c+\dotsb,\\
\nu^i_j(\pi)&=\delta^i_j+f^i_{aj}\pi^a+\frac12f^i_{ak}f^k_{bj}\pi^a\pi^b+\dotsb.
\end{split}
\label{expparam}
\end{equation}
This allows one to work out explicitly both the kinetic terms and interactions of NG bosons. 

For various practical purposes, it is also useful to have an explicit expression for the symmetry transformation of the NG fields. This is discussed in detail below in section~\ref{subsec:EFTsymmetries}; its finite and infinitesimal versions read
\begin{equation}
U(\pi')=\g U(\pi)\h(\pi,\g)^{-1}=e^{\imag\epsilon^iT_i}U(\pi)e^{-\imag\epsilon^ik^\alpha_i(\pi)T_\alpha},
\label{defk}
\end{equation}
where $\g=e^{\imag\epsilon^iT_i}\in G$ and $\h\in H$. The infinitesimal shift of the NG fields is denoted as $\delta\pi^a=\epsilon^ih^a_i(\pi)$. In geometrical terms, the functions $h^a_i(\pi)$ define infinitesimal group motions on the coset space $G/H$, and thus correspond to the Killing vectors of the symmetry group $G$. Multiplying eq.~(\ref{defk}) from the left by $U(\pi)^{-1}$ and expanding to first order in $\epsilon^i$, we obtain the simple relations $\nu^a_i=\omega^a_bh^b_i$ and $\nu^\alpha_i=\omega^\alpha_ah^a_i+k^\alpha_i$. Using the already known expressions for $\omega^i_a$ and $\nu^i_j$, we can solve these equations iteratively, and obtain for the exponential parameterization $U(\pi)=e^{\imag\pi^aT_a}$~\cite{Watanabe:2014fva}
\begin{equation}
h^a_i(\pi)=\delta^a_i-\Bigl(f^a_{ib}+\frac12f^a_{bc}\delta^c_i\Bigr)\pi^b+\dotsb,\qquad k^\alpha_i(\pi)=\delta^\alpha_i-\Bigl(f^\alpha_{ib}+\frac12f^\alpha_{bc}\delta^c_i\Bigr)\pi^b+\dotsb.
\end{equation}
Furthermore, we can now give a particularly simple interpretation for the auxiliary field $\phi^a_\mu$. Plugging the relation $\nu^a_i=\omega^a_bh^b_i$ into eq.~\eqref{auxfields}, this field can namely be written as
\begin{equation}
\phi^a_\mu(\pi)=-\omega^a_b(\pi)D_\mu\pi^b,
\label{NGcovder}
\end{equation}
where $D_\mu\pi^a=\de_\mu\pi^a-A^i_\mu h^a_i(\pi)$ is a covariant derivative of the NG field. Note that this agrees with the usual notion of a covariant derivative: the coefficient $h^a_i(\pi)$ in front of $A^i_\mu$ defines an infinitesimal symmetry transformation of the field $\pi^a$.


\section{Examples}
\label{sec:examples}

\subsection{Symmetric coset spaces --- general considerations}

In many cases of physical interest the coset space $G/H$ turns out to be \emph{symmetric}. This means that the commutator of two broken generators is a linear combination of unbroken generators only, or $f^a_{bc}=0$. Formally, this property is equivalent to the existence of an automorphism $R$ of the Lie algebra of $G$, under which $R(T_\alpha)=T_\alpha$ and $R(T_a)=-T_a$. Choosing the parameterization $U(\pi)=e^{\imag\pi^aT_a}$ and applying the automorphism $R$ to the transformation rule of eq.~\eqref{defk} gives $R(U')=U'^{-1}=R(\g)U^{-1}\h^{-1}$. Taking the inverse of this expression and multiplying it with eq.~\eqref{defk}, we infer that there is a field variable which, unlike $U(\pi)$, transforms linearly under the entire group $G$,
\begin{equation}
\Sigma(\pi)=U(\pi)^2,\qquad
\Sigma(\pi')=\g\Sigma(\pi)R(\g)^{-1}.
\label{defSigma}
\end{equation}
Due to this property, $\Sigma(\pi)$ (or an equivalent variable) is often taken as the starting point of the construction of EFTs. We should nevertheless emphasize that $\phi^a_\mu$ and $B^\alpha_\mu$ are conceptually more convenient, as they carry a derivative, implying that when expressed in terms of them, the effective Lagrangian contains only a finite number of contributions at every order in the derivative expansion. At the same time, adding a factor of $\Sigma$ or $\Sigma^{-1}$ does not increase the order of a given operator, and one thus has to go through some extra effort to classify all the possible terms in the Lagrangian.

The advantage of the notation~\eqref{defSigma} is, however, that it makes it trivial to construct the covariant derivative
\begin{equation}
D_\mu\Sigma=\de_\mu\Sigma-\imag A_\mu\Sigma+\imag\Sigma R(A_\mu),
\end{equation}
as well as to take higher derivatives. Applying the automorphism $R$ to the definition of our auxiliary fields~\eqref{Bphidefprel}, we can project out the broken part and show that it equals
\begin{equation}
\phi_\mu=+\frac\imag2U^{-1}(D_\mu\Sigma)U^{-1}=-\frac\imag2U(D_\mu\Sigma^{-1})U.
\label{Sigma}
\end{equation}
Upon a straightforward although somewhat lengthy manipulation, a similar expression can be found for $D_\mu\phi_\nu$; one possible and rather convenient formulation for it is
\begin{equation}
D_\mu\phi_\nu=\frac\imag4\bigl[U^{-1}(D_\mu D_\nu\Sigma)U^{-1}-U(D_\mu D_\nu\Sigma^{-1})U\bigr].
\label{DSigma}
\end{equation}
To complete the dictionary between the two formalisms, we still need to find an expression for $G^\alpha_{\mu\nu}$ in terms of linearly transforming variables. To this end, recall that a field-strength tensor transforms covariantly, and hence by eq.~\eqref{Bphidefprel} the field-strength tensor of the original gauge field $A^i_\mu$ is related to one expressed in terms of $\phi^a_\mu$ and $B^\alpha_\mu$ via
\begin{align}
\notag
U^{-1}F_{\mu\nu}U&=\de_\mu B_\nu-\de_\nu B_\mu-\imag[B_\mu,B_\nu]-\imag[\phi_\mu,\phi_\nu]+\de_\mu\phi_\nu-\de_\nu\phi_\mu-\imag[B_\mu,\phi_\nu]+\imag[B_\nu,\phi_\mu]\\
&=G_{\mu\nu}-\imag[\phi_\mu,\phi_\nu]+D_\mu\phi_\nu-D_\nu\phi_\mu.
\label{Fconj}
\end{align}
This allows us to express $G_{\mu\nu}$ in terms of $\phi_\mu$, $D_\mu\phi_\nu$, and $F_{\mu\nu}$, of which the former two are given above in eqs.~\eqref{Sigma} and~\eqref{DSigma}.


\subsection{Pions in quantum chromodynamics}
\label{subsec:pions}

QCD possesses, apart from spacetime Poincar\'e invariance, an approximate global $\gr{SU}(N)_\text{L}\times\gr{SU}(N)_\text{R}$ symmetry under independent unitary transformations of left- and right-handed quarks, where $N$ is the number of light quark flavors. The physically relevant cases are $N=2,3$. In the ground state, this chiral symmetry is spontaneously broken to its diagonal subgroup, $H=\gr{SU}(N)_\text{V}$, which leads to the spectrum of QCD containing $N^2-1$ light pseudo-NG bosons, denoted here collectively as pions. The low-energy EFT for pions (and possibly other, heavier degrees of freedom) is the celebrated $\chi$PT, originally developed in refs.~\cite{Gasser:1983yg,Gasser:1984gg}.


\subsubsection{Coset fields and symmetry transformations}

It is customary to represent the direct product structure of the chiral group using brackets; a general element of the group takes the form $\cpr{\g_L}{\g_R}$ with $\g_L,\g_R\in\gr{SU}(N)$. The unbroken subgroup corresponds to elements of the type $\cpr\g\g$ and is generated by a linear combination of the left and right generators, $\cpr T\openone+\cpr\openone T$. The broken generators can be chosen orthogonal, $\cpr T\openone-\cpr\openone T$, and the coset element thus reads $U=\cpr u{u^{-1}}$. The transformation rule~\eqref{defk} reads accordingly $\cpr{u'}{{u'}^{-1}}=\cpr{\g_L}{\g_R}\cpr u{u^{-1}}\cpr{\h^{-1}}{\h^{-1}}$. The coset space is symmetric due to the automorphism acting on the group as $R\cpr{\g_L}{\g_R}=\cpr{\g_R}{\g_L}$. The linearly transforming variable $\Sigma=U^2=\cpr{u^2}{u^{-2}}$, see eq.~\eqref{defSigma}, can be traded for the matrix $\CPT=u^2$ that transforms as $\CPT'=\g_L\CPT\g_R^{-1}$; this is the field variable that is usually used to construct the Lagrangian of $\chi$PT. Each of the $\gr{SU}(N)$ subgroups is associated with an independent set of gauge fields, in terms of which the total matrix gauge field reads $A_\mu=\cpr{A^L_\mu}\openone+\cpr\openone{A^R_\mu}$. The covariant derivative of $\Sigma$ then decomposes as $D_\mu\Sigma=(D_\mu\CPT,\CPT^{-1})+(\CPT,D_\mu\CPT^{-1})$, where
\begin{equation}
D_\mu\CPT=\de_\mu\CPT-\imag A^L_\mu\CPT+\imag\CPT A^R_\mu.
\end{equation}
Likewise, eq.~\eqref{Sigma} becomes
\begin{equation}
\phi_\mu=\cpr{\Phi_\mu}\openone-\cpr\openone{\Phi_\mu},\qquad
\pCPT_\mu=+\frac\imag2u^{-1}(D_\mu\CPT)u^{-1}=-\frac\imag2u(D_\mu\CPT^{-1})u.
\label{Phidef}
\end{equation}
Finally, carrying out the conjugation of the field-strength tensor indicated in eq.~\eqref{Fconj} and projecting out the unbroken part of the result with the help of the automorphism $R$, we obtain the relation
\begin{equation}
\BCPT_{\mu\nu}=\frac12(u^{-1}F^L_{\mu\nu}u+uF^R_{\mu\nu}u^{-1})+\imag[\pCPT_\mu,\pCPT_\nu],
\label{uFu}
\end{equation}
where the Lie algebra valued field $\BCPT_{\mu\nu}$ is defined naturally by $G_{\mu\nu}=\cpr{\BCPT_{\mu\nu}}\openone+\cpr\openone{\BCPT_{\mu\nu}}$.

Since for the symmetry-breaking pattern of $\chi$PT both broken and unbroken generators transform in the adjoint representation of $H$, the matrix elements of both basic building blocks, $(\pCPT_\mu)^A_B$ and $(\BCPT_{\mu\nu})^A_B$, as well as of their covariant derivatives constitute a traceless tensor of $H$ with the upper index $A$ transforming in the fundamental representation and the lower index $B$ in its complex conjugate. The fundamental representation of $\gr{SU}(N)$ has three algebraically independent invariant tensors, namely $\delta^A_B$, $\epsilon^{ABC\dotsb}$ and $\epsilon_{ABC\dotsb}$;\footnote{This can be viewed as a consequence of the definition of $\gr{SU}(N)$ as the set of all complex $N\times N$ matrices satisfying the conditions $UU^\dagger=\openone$ and $\det U=1$, which precisely encode the invariance of $\delta^A_B$ and $\epsilon_{ABC\dotsb}$. The absence of any other algebraically independent invariant tensor means that the matrices do not satisfy any other independent algebraic constraints~\cite{Georgi:1982jb}.} every term in the invariant Lagrangian can be obtained by contracting the indices of $\Phi_\mu$ and $\BCPT_{\mu\nu}$ (and possibly their covariant derivatives) with products of these tensors. Moreover, since all our fields have the same number of upper and lower indices, such an invariant term must necessarily contain the same number of $\epsilon^{ABC\dotsb}$ and $\epsilon_{ABC\dotsb}$, and can therefore be decomposed into products of $\delta^A_B$ alone. In short, \emph{every invariant term in the Lagrangian can be written as a product of traces of $\pCPT_\mu$, $\BCPT_{\mu\nu}$ and their covariant derivatives}.


\subsubsection{Invariant Lagrangians}

At the leading order ($s+t=2$), there is only one possible operator that can be assembled from the available building blocks applying the strategy described above; using eq.~\eqref{Phidef}, the invariant Lagrangian thus acquires the form
\begin{equation}
\La^{(2)}_\text{inv}\propto\tr(\pCPT_\mu\pCPT^\mu)=\frac14\tr(D_\mu\CPT D^\mu\CPT^{-1})
\end{equation}
up to an overall factor that defines the pion decay constant. This agrees with the fact that the NG bosons span an irreducible multiplet of $H$.

At the next-to-leading order ($s+t=4$), the list of possible operators in the invariant Lagrangian is considerably longer, see section~\ref{subsubsec:LLI}. Taking into account the fact that operators of the type $\epsilon^{\kappa\lambda\mu\nu}\phi^a_\kappa\phi^b_\lambda\phi^c_\mu\phi^d_\nu$ and $\epsilon^{\kappa\lambda\mu\nu}\phi^a_\kappa\phi^b_\lambda D_\mu\phi^c_\nu$ do not contribute due to the cyclicity of the trace (the former vanishes at the level of the Lagrangian, while the latter evaluates to a mere surface term), we obtain for the order-four invariant Lagrangian 
\begin{align}
\notag
\La^{(4)}_\text{inv}={}&c_1\tr(\pCPT_\mu\pCPT^\mu\pCPT_\nu\pCPT^\nu)+c_2\tr(\pCPT_\mu\pCPT_\nu\pCPT^\mu\pCPT^\nu)+c_3\tr(\pCPT_\mu\pCPT^\mu)\tr(\pCPT_\nu\pCPT^\nu)\\
\label{order4naive}
&+c_4\tr(\pCPT_\mu\pCPT_\nu)\tr(\pCPT^\mu\pCPT^\nu)+c_5\tr(\pCPT^\mu\pCPT^\nu D_\mu\pCPT_\nu)+c_6\tr(\pCPT^\nu\pCPT^\mu D_\mu\pCPT_\nu)\\
\notag
&+c_7\tr(D_\mu\pCPT_\nu D^\mu\pCPT^\nu)+c_8\tr(D_\mu\pCPT^\mu D_\nu\pCPT^\nu)+c_{9}\tr(\pCPT^\mu\pCPT^\nu \BCPT_{\mu\nu})+c_{10}\tr(D^\mu\pCPT^\nu \BCPT_{\mu\nu})\\
\notag
&+c_{11}\tr(\BCPT_{\mu\nu}\BCPT^{\mu\nu})+c_{12}\epsilon^{\kappa\lambda\mu\nu}\tr(\pCPT_\kappa\pCPT_\lambda \BCPT_{\mu\nu}).
\end{align}
In deriving this result, we have only used the invariance of the Lagrangian under the continuous $\gr{SU}(N)_\text{L}\times\gr{SU}(N)_\text{R}$ symmetry. Nevertheless, QCD is in addition invariant under the discrete symmetries of parity, charge conjugation and time reversal. We may use the fact that under parity, the pion fields transform as $\pi^a(x)\to-\pi^a(\mathcal P x)$, where $\mathcal P^\mu_{\phantom\mu\nu}=\text{diag}(1,-1,-1,-1)$ is the spatial inversion matrix, in addition to which the left- and right-handed background gauge fields are interchanged. In our notation, the parity transformation can be expressed compactly as~\cite{Scherer:2002tk}
\begin{equation}
\phi_\mu(x)\to-\mathcal P_\mu^{\phantom\mu\nu}\phi_\nu(\mathcal P x),\qquad B_\mu(x)\to\mathcal P_\mu^{\phantom\mu\nu}B_\nu(\mathcal P x).
\end{equation}
Parity invariance of QCD thus directly rules out the $c_5$, $c_6$, $c_{10}$ and $c_{12}$ operators.

The number of independent operators in the Lagrangian~\eqref{order4naive} can be further reduced by using the special algebraic properties of traceless matrices of dimension $N=2,3$~\cite{Fearing:1994ga}. In both cases the identity $\tr(X^4)=\frac12[\tr(X^2)]^2$ holds. Substituting $X=aA+bB+cC+dD$ where $A,B,C,D$ are traceless matrices and $a,b,c,d$ numerical coefficients and comparing the terms proportional to $abcd$, one obtains
\begin{equation}
\begin{split}
&\tr(ABCD+ABDC+ACBD+ACDB+ADBC+ADCB)\\
&=\tr(AB)\tr(CD)+\tr(AC)\tr(BD)+\tr(AD)\tr(BC).
\end{split}
\label{ABCDidentity}
\end{equation}
This in turn leads to a relation among the $c_1$, $c_2$, $c_3$ and $c_4$ operators,
\begin{equation}
\tr(4\pCPT_\mu\pCPT^\mu\pCPT_\nu\pCPT^\nu+2\pCPT_\mu\pCPT_\nu\pCPT^\mu\pCPT^\nu)=\tr(\pCPT_\mu\pCPT^\mu)\tr(\pCPT_\nu\pCPT^\nu)+2\tr(\pCPT_\mu\pCPT_\nu)\tr(\pCPT^\mu\pCPT^\nu),
\end{equation}
which allows us to eliminate one of them, say $c_2$. In fact, for $N=2$ the trace of a product of four generators can be resolved in terms of traces of a product of two generators only using the special properties of Pauli matrices, leading to
\begin{equation}
\begin{split}
\tr(\pCPT_\mu\pCPT^\mu\pCPT_\nu\pCPT^\nu)&=\frac12\tr(\pCPT_\mu\pCPT^\mu)\tr(\pCPT_\nu\pCPT^\nu),\\
\tr(\pCPT_\mu\pCPT_\nu\pCPT^\mu\pCPT^\nu)&=\tr(\pCPT_\mu\pCPT_\nu)\tr(\pCPT^\mu\pCPT^\nu)-\frac12\tr(\pCPT_\mu\pCPT^\mu)\tr(\pCPT_\nu\pCPT^\nu).
\end{split}
\end{equation}
Hence for $N=2$, both $c_1$ and $c_2$ can be eliminated. All the independent operators in this set ($c_3$ and $c_4$ for $N=2$, and $c_1$, $c_3$ and $c_4$ for $N=3$) can be easily rewritten in terms of $\CPT$ using eq.~\eqref{Phidef}.

What remains to be discussed are possible redundancies among the operators $c_7$, $c_8$, $c_9$ and $c_{11}$. As elaborated on in section~\ref{subsec:EOM}, the leading-order equation of motion can be used to simplify the effective Lagrangian at higher orders and thereby to reduce the number of independent effective coupling constants. In Lorentz-invariant systems with a symmetric coset space, this equation of motion reduces to $D_\mu\phi^\mu=0$; see eq.~\eqref{EOMLO}. The $c_8$ operator is therefore redundant. Furthermore, the $c_9$ and $c_{11}$ operators can be expressed in terms of the physical field-strength tensors $F^{L,R}_{\mu\nu}$ by means of eq.~\eqref{uFu}. Since a trace of four factors of $\Phi_\mu$ is already present in the $c_1$ and $c_2$ terms, this gives us two new operators,
\begin{equation}
\begin{split}
\tr(u^{-1}F^L_{\mu\nu}u+uF^R_{\mu\nu}u^{-1})^2&=\tr(F^L_{\mu\nu}F^{L\mu\nu}+F^R_{\mu\nu}F^{R\mu\nu}+2F^L_{\mu\nu}\CPT F^{R\mu\nu}\CPT^{-1}),\\
\tr\{(u^{-1}F^L_{\mu\nu}u+uF^R_{\mu\nu}u^{-1})[\Phi^\mu,\Phi^\nu]\}&=\frac12\tr(F^L_{\mu\nu}D^\mu\CPT D^\nu\CPT^{-1}+F^R_{\mu\nu}D^\mu\CPT^{-1}D^\nu\CPT).
\end{split}
\label{c9c11}
\end{equation}
Finally, one can show that the $c_7$ operator gives, up to terms that vanish due to equation of motion, an expression identical to the first line of eq.~\eqref{c9c11}, just with an opposite sign in front of $2F^L_{\mu\nu}\CPT F^{R\mu\nu}\CPT^{-1}$. Since to see this requires some effort, we present the details in appendix~\ref{app:c7} in order not to interrupt the flow of the argument here.

Altogether, choosing a suitable basis of operators and redefining the coupling constants appropriately, the most general invariant Lagrangian for QCD with two or three light quark flavors at order four in derivatives acquires the form
\begin{equation}
\begin{split}
\La^{(4)}_\text{inv}={}&\tilde{c}_1\tr(D_\mu\CPT D^\mu\CPT^{-1}D_\nu\CPT D^\nu\CPT^{-1})+\tilde{c}_2\tr(D_\mu\CPT D^\mu\CPT^{-1})\tr(D_\nu\CPT D^\nu\CPT^{-1})\\
&+\tilde{c}_3\tr(D_\mu\CPT D_\nu\CPT^{-1})\tr(D^\mu\CPT D^\nu\CPT^{-1})\\
&+\tilde c_4\tr(F^L_{\mu\nu}D^\mu\CPT D^\nu\CPT^{-1}+F^R_{\mu\nu}D^\mu\CPT^{-1}D^\nu\CPT)+\tilde c_5\tr(F^L_{\mu\nu}\CPT F^{R\mu\nu}\CPT^{-1})\\
&+\tilde c_6\tr(F^L_{\mu\nu}F^{L\mu\nu}+F^R_{\mu\nu}F^{R\mu\nu}).
\end{split}
\label{order4final}
\end{equation}
Up to possible difference in notation, this is recognized as the familiar order-four Lagrangian of $\chi$PT; the $\tilde c_1$ term is redundant in the $N=2$ case~\cite{Scherer:2002tk}.


\subsubsection{Explicit symmetry breaking}

In real QCD, invariance of the Lagrangian under the full chiral group is violated in a twofold manner. First, at the fourth order in derivatives, the effects of the chiral anomaly, whose discussion goes beyond the scope of the present paper, enter the game~\cite{Witten:1983tw}. Second, the chiral symmetry is broken explicitly by nonzero quark masses. These appear in the microscopic Lagrangian of QCD through the mass term $\bar\psi_L\mathcal M\psi_R+\bar\psi_R\mathcal M^\dagger\psi_L$, where $\mathcal M$ is the quark mass matrix. It is real and diagonal, yet we treat it as a complex matrix that transforms under a chiral rotation as $\mathcal M\to\g_L\mathcal M\g_R^{-1}$. One can think of $\mathcal M$ as a background (pseudo)scalar field akin to $A_\mu$: gauge invariance restricts the way that $\mathcal M$ appears in the low-energy EFT, and only at the end of the day one sets $\mathcal M=\text{diag}(m_u,m_d,m_s)$. In line with our general procedure, the effective Lagrangian will be expressed in terms of the composite field $\Xi=u^{-1}\mathcal Mu^{-1}$ that transforms as a complex adjoint field plus a complex singlet (corresponding to $\tr\Xi$) of the unbroken subgroup. Note that under parity, $\Xi(x)\to\Xi(\mathcal Px)^\dagger$; this further constrains the way that $\Xi$ can appear in the Lagrangian.

At the lowest, second order in derivatives, there is only one chirally invariant operator preserving parity, given solely by the singlet part of $\Xi$,
\begin{equation}
\La_\text{s.b.}^{(0)}\propto\tr(\Xi+\Xi^\dagger)=\tr(\mathcal M\CPT^\dagger+\mathcal M^\dagger\CPT),
\end{equation}
up to an overall factor that is to be treated as a free parameter. At the fourth order, the operators that contribute can be read off the list provided in section~\ref{subsubsec:LLI}: $\Xi_\rho\Xi_\sigma$, $\Xi_\sigma\phi^a_\mu\phi^{b\mu}$ and $\Xi_\sigma D_\mu\phi^{a\mu}$. The latter can be eliminated by using the equation of motion~\eqref{EOMLO}. With the additional constraint due to parity, the remaining two operators give the following,
\begin{equation}
\begin{split}
\La_\text{s.b.}^{(2)}={}&d_1\tr(\mathcal M\CPT^\dagger)\tr(\mathcal M^\dagger\CPT)+d_2[(\tr\mathcal M\CPT^\dagger)^2+(\tr\mathcal M^\dagger\CPT)^2]\\
&+d_3\tr(\mathcal M\CPT^{\dagger}\mathcal M\CPT^{\dagger}+\mathcal M^\dagger\CPT\mathcal M^\dagger\CPT)+d_4\tr(\mathcal M\mathcal M^\dagger)\\
&+d_5\tr(\mathcal M\CPT^\dagger+\mathcal M^\dagger\CPT)\tr(D_\mu\CPT D^\mu\CPT^\dagger)+d_6\tr[(\mathcal M\CPT^\dagger+\CPT\mathcal M^\dagger)D_\mu\CPT D^\mu\CPT^\dagger].
\end{split}
\end{equation}
For $N=2$, we have an identity similar to eq.~\eqref{ABCDidentity},
\begin{equation}
\tr(ABC+BAC)=\tr(AB)\tr(C),
\end{equation}
valid for traceless $A$, $B$ and arbitrary $C$. (The proof is trivial in case $A$ and $B$ are Hermitian and therefore inherit the anticommutation properties of Pauli matrices.) Applying this to $\tr(\Xi\pCPT_\mu\pCPT^\mu)$ reveals that the $d_6$ operator can be expressed in terms of the $d_5$ one for $N=2$.


\subsection{Spin waves in ferromagnets}
\label{subsec:magnons}

Ferromagnets are nonrelativistic systems with a global \emph{internal} $G=\gr{SU}(2)$ spin symmetry, which is broken by the spontaneous magnetization in the ground state to its $H=\gr{U}(1)$ subgroup. The two broken generators correspond to one NG mode in the spectrum: the spin wave, or magnon. Since its dispersion relation is quadratic at low momentum, the derivative expansion of the effective Lagrangian has to be organized accordingly; see section~\ref{subsec:powercounting} for more details. One temporal derivative counts as two spatial ones~\cite{Leutwyler:1993gf}, as a result of which up to order four in momenta, only $\La_\text{eff}^{(1,0)}$, $\La_\text{eff}^{(0,1)}$, $\La_\text{eff}^{(2,0)}$, $\La_\text{eff}^{(1,1)}$, $\La_\text{eff}^{(3,0)}$, $\La_\text{eff}^{(0,2)}$, $\La_\text{eff}^{(2,1)}$ and $\La_\text{eff}^{(4,0)}$ need to be taken into account. In order to simplify our discussion, we will restrict ourselves to isotropic (rotationally invariant) ferromagnets in three spatial dimensions. This immediately rules out all operators from $\La_\text{eff}^{(1,0)}$ and $\La_\text{eff}^{(1,1)}$. Moreover, we will assume that the system is invariant under parity. Since angular momentum is an axial vector, parity is not spontaneously broken in the ground state and the NG fields are parity-even: $\pi^a(x)\to\pi^a(\mathcal Px)$. As a consequence, $\La_\text{eff}^{(3,0)}$ is ruled out.


\subsubsection{Leading-order Lagrangian}

The leading, order-two Lagrangian is given by two pieces, $\La_\text{eff}^\text{LO}=\La_\text{eff}^{(0,1)}+\La_\text{eff}^{(2,0)}$. According to sections~\ref{subsubsec:Linvar} and~\ref{subsubsec:LCS}, the available operators are $e_\alpha B^\alpha_0$, $e_a\phi^a_0$ and $\bar g_{ab}\phi^a_r\phi^b_r$. Note that $e_a$ vanishes, being equal to the density of broken generators in the ground state. The coset space $\gr{SU(2)/U(1)}$ is symmetric; choosing the magnetization of the ground state without loss of generality to point in the third spin direction, the corresponding automorphism can be realized using the third Pauli matrix, $R(\g)=\sigma_3\g\sigma_3$. One can then trade the linearly transforming variable $\Sigma$ of eq.~\eqref{defSigma} for
\begin{equation}
N(\pi)=\Sigma(\pi)\sigma_3=U(\pi)\sigma_3U(\pi)^{-1}.
\end{equation}
The matrix $N$ transforms in the adjoint representation of $G$, $N(\pi')=\g N(\pi)\g^{-1}$. Being traceless, Hermitian and involutory, it can be equivalently expressed in terms of a unit vector $\vec n(\pi)$ as $N=\vec n\cdot\vec\sigma$. It is this variable that is usually used to write down the EFT for ferromagnets. The invariant part of the leading-order Lagrangian then reads
\begin{equation}
\La_\text{eff}^{(2,0)}=-\frac{\rho_s}4\tr(D_rND_rN)=-\frac{\rho_s}2D_r\vec n\cdot D_r\vec n,
\end{equation}
where $D_\mu\vec n=\de_\mu\vec n+\vec A_\mu\times\vec n$ is the covariant derivative. The gauge potentials $\vec A_\mu$ can be interpreted in terms of the intensities of external electric and magnetic fields~\cite{Frohlich:1993gs}, and the parameter $\rho_s$ is usually referred to as the spin stiffness. 

Unlike $\La^{(2,0)}_\text{eff}$, the CS part of the Lagrangian, $\La^{(0,1)}_\text{eff}=e_\alpha B^\alpha_0$, cannot be written in a manifestly invariant form in terms of $\vec n$~\cite{Leutwyler:1993gf,Roman:1999ro}. There are several different, physically equivalent but mathematically distinct, expressions for it. The most straightforward one is based on a mere power expansion in the NG fields using eq.~\eqref{expparam}. It is, in fact, possible to write the Lagrangian solely in terms of $\vec n$, but only at the cost of extending the spacetime by one extra dimension~\cite{Volovik:1987li,Leutwyler:1993gf,CSpaper}. This way, one can derive the expression
\begin{equation}
\La_\text{eff}^{(0,1)}=m\frac{\dot n_1n_2-\dot n_2n_1}{1+n_3}+m\vec A_0\cdot\vec n;
\label{ferroL01}
\end{equation}
the effective coupling $m$ has the interpretation of the spin density in the ground state. The temporal field $\vec A_0$ stands, up to a factor, for the external magnetic field intensity, and the term $m\vec A_0\cdot\vec n$ therefore represents the usual Zeeman coupling of spin.


\subsubsection{Next-to-leading-order Lagrangian}

The next contributions to the Lagrangian, $\La_\text{eff}^{(0,2)}$, $\La_\text{eff}^{(2,1)}$ and $\La_\text{eff}^{(4,0)}$, are of order four in momenta. Up to an overall factor and the replacement $D_r\to D_0$, the piece $\La_\text{eff}^{(0,2)}$ is identical to $\La_\text{eff}^{(2,0)}$~\cite{Leutwyler:1993gf}. Moreover, being bilinear in $\phi^a_0$ it is actually irrelevant, for it can be eliminated using the equation of motion~\eqref{EOMLO}, which is linear in $\phi^a_0$ when $\La_\text{eff}^{(0,1)}$ is present. As to $\La^{(2,1)}_\text{eff}$, four different operators are available in three spatial dimensions; see the list in section~\ref{subsubsec:Linvar}. Out of these, $\phi^a_0\phi^b_r\phi^c_r$ and $\phi^a_rG^\alpha_{0r}$ are clearly forbidden by symmetry since $\phi^a_\mu$ transforms as a two-vector of the unbroken $\gr{U(1)\simeq SO(2)}$ whereas $B^\alpha_\mu$ is a singlet. In addition, the operator $\phi^a_0D_r\phi^b_r$ can again be eliminated by using the leading-order equation of motion~\eqref{EOMLO}. Altogether, only one type of operator is therefore present: $\phi^a_rD_0\phi^b_r$. In order to make it invariant, the spin indices must be contracted either with $\delta_{ab}$ or with $\epsilon_{ab}$. The former however leads to an operator that is a total time derivative so that only the latter can give a nontrivial result. Using eqs.~\eqref{Sigma} and~\eqref{DSigma} as well as the fact that in terms of matrices, $\epsilon_{ab}\phi^b_\mu$ is proportional to $[\sigma_3,\phi_\mu]$, we obtain upon a short manipulation
\begin{equation}
\La_\text{eff}^{(2,1)}\propto\tr([\sigma_3,\phi_r]D_0\phi_r)\propto(\vec n\times D_r\vec n)\cdot D_0D_r\vec n.
\label{ferroL21}
\end{equation}
Note that despite containing just one time derivative, this interaction is strictly invariant under time reversal, for this transforms the spin vector as $\vec n(t,\bm x)\to-\vec n(-t,\bm x)$. Under the same transformation, the Lagrangian~\eqref{ferroL01} shifts by a total time derivative.

Let us finally construct $\La_\text{eff}^{(4,0)}$. Here we have seven different operators in three spatial dimensions, two of which ($\phi^a_r\phi^b_sD_r\phi^c_s$ and $D_r\phi^a_sG^\alpha_{rs}$) are immediately seen to vanish by the unbroken $\gr{SO(2)}$ symmetry.  We shall consider the remaining operators in the order given in section~\ref{subsubsec:Linvar}. In order to see how to combine the indices in $\phi^a_r\phi^b_r\phi^c_s\phi^d_s$ so as to get an invariant, it is suitable to think of the two components of $\phi^a_\mu$ as the real and imaginary parts of a complex field $\Phi_\mu$. Under the unbroken $\gr{U(1)}$ symmetry, this acquires a phase. It is now obvious that there are two independent invariant operators, $\Phi^*_r\Phi_r\Phi^*_s\Phi_s$ and $\Phi^*_r\Phi^*_r\Phi_s\Phi_s$, which can be mapped to linear combinations of $\tr(\phi_r\phi_r)\tr(\phi_s\phi_s)$ and $\tr(\phi_r\phi_s)\tr(\phi_r\phi_s)$. In terms of the unit vector $\vec n$, these can be rewritten as
\begin{equation}
\La_\text{eff}^{(4,0)}\supset e_1(D_r\vec n\cdot D_r\vec n)(D_s\vec n\cdot D_s\vec n)+e_2(D_r\vec n\cdot D_s\vec n)(D_r\vec n\cdot D_s\vec n).
\end{equation}
The operators of the type $D_r\phi^a_sD_r\phi^b_s$ and $D_r\phi^a_rD_s\phi^b_s$ are straightforward to evaluate using eq.~\eqref{DSigma} and the trace properties of Pauli matrices. One thus finds, for instance, $D_r\phi^a_sD_r\phi^b_s\propto(D_rD_s\vec n)^2-(\vec n\cdot D_rD_s\vec n)^2=(D_rD_s\vec n)^2-(D_r\vec n\cdot D_s\vec n)^2$, where we used the fact that $\vec n\cdot D_s\vec n=0$. The last term is already contained in the $e_2$ operator. Altogether, the Lagrangian therefore acquires two new independent operators,
\begin{equation}
\La_\text{eff}^{(4,0)}\supset e_3D_rD_r\vec n\cdot D_sD_s\vec n+\tilde e_3D_rD_s\vec n\cdot D_rD_s\vec n.
\end{equation}
The remaining two types of operators, $\phi^a_r\phi^b_sG^\alpha_{rs}$ and $G^\alpha_{rs}G^\beta_{rs}$, both contain the auxiliary gauge field $G^\alpha_{\mu\nu}$. In ferromagnets this has only one component, and is found with the help of eq.~\eqref{Fconj} to be
\begin{equation}
G^3_{rs}\propto \vec n\cdot\vec F_{rs}-\vec n\cdot(D_r\vec n\times D_s\vec n).
\end{equation}
The second term arises from the bilinear $[\phi_r,\phi_s]$ and its square is already contained in the operators $e_1$ and $e_2$. Using finally the fact that $\vec n\cdot D_\mu\vec n=0$ and thus $D_r\vec n\times D_s\vec n$ is parallel to $\vec n$, we can write the two new operators contributing to the Lagrangian as
\begin{equation}
\La_\text{eff}^{(4,0)}\supset e_4\vec F_{rs}\cdot(D_r\vec n\times D_s\vec n)+e_5(\vec n\cdot\vec F_{rs})(\vec n\cdot\vec F_{rs}).
\end{equation}

The above-found operators already span a basis that gives the most general effective Lagrangian at order four in momenta compatible with the symmetry. However, it is convenient to switch to a somewhat different basis in which the dependence on the NG and background fields is more transparent. The argument closely resembles the one in appendix~\ref{app:c7} by which the operator $c_7$ is eliminated from the $\chi$PT Lagrangian. Namely, using integration by parts and the fact that the commutator of covariant derivatives $[D_r,D_s]$ is proportional to $F_{rs}$, the $\tilde e_3$ operator is found to be a linear combination of the $e_3$, $e_4$, $e_5$ ones and of $\vec F_{rs}\cdot\vec F_{rs}$.

The operator in $\La_\text{eff}^{(2,1)}$ can be handled in the same way. We first write $D_0D_r\vec n=[D_0,D_r]\vec n+D_rD_0\vec n$ and observe that the first term leads to an operator of the type $\vec F_{0r}\cdot D_r\vec n$. As to the second term, note that the leading-order equation of motion~\eqref{EOMLO} takes for ferromagnets the form $mD_0\vec n=\rho_s\vec n\times D_rD_r\vec n$~\cite{Leutwyler:1993gf}. Hence $(\vec n\times D_r\vec n)\cdot D_rD_0\vec n$ can be absorbed into a redefinition of the couplings $e_1$ and $e_3$. Putting all the pieces together, we then obtain the most general effective Lagrangian for an isotropic, parity-invariant ferromagnet up to order four in momenta, which we collect here for the reader's sake,
\begin{equation}
\begin{split}
\La_\text{eff}={}&m\frac{\dot n_1n_2-\dot n_2n_1}{1+n_3}+m\vec A_0\cdot\vec n-\frac{\rho_s}2D_r\vec n\cdot D_r\vec n\\
&+e_1(D_r\vec n\cdot D_r\vec n)(D_s\vec n\cdot D_s\vec n)+e_2(D_r\vec n\cdot D_s\vec n)(D_r\vec n\cdot D_s\vec n)\\
&+e_3D_rD_r\vec n\cdot D_sD_s\vec n+e_4\vec F_{rs}\cdot(D_r\vec n\times D_s\vec n)+e_5(\vec n\cdot\vec F_{rs})(\vec n\cdot\vec F_{rs})\\
&+e_6\vec F_{rs}\cdot\vec F_{rs}+e_7\vec F_{0r}\cdot D_r\vec n.
\end{split}
\end{equation}
The order-four part of the Lagrangian contains seven independent couplings. In the literature, a somewhat reduced Lagrangian (see, for instance, ref.~\cite{Hofmann:2001ck}) is usually employed which can be obtained as follows. In the absence of external electric fields, $\vec A_r=\vec0$, hence the operators $e_4$, $e_5$ and $e_6$ disappear. If in addition the background magnetic field is uniform, then $\vec F_{0r}=\vec 0$ and the $e_7$ operator drops out as well. In uniform magnetic fields, the order-four Lagrangian thus contains only three independent couplings: $e_1$, $e_2$ and $e_3$.


\section{Methodology}
\label{sec:method}

The problem of constructing the EFT can be transformed into an elementary exercise in field theory by following a number of straightforward intermediate steps. In order to stress the importance of these steps, and because they can be easily discussed on their own footing, we formulate some of them as standalone ``theorems''. Their proofs are either well known or can be found in the literature, and we therefore only show details where it helps to clarify the argument.


\subsection{Symmetries of the effective theory}
\label{subsec:EFTsymmetries}

Consider now a system with a continuous internal symmetry group $G$. Each independent generator $T_i$ of this group gives rise to a conserved Noether current. When the ground state of the system breaks the symmetry spontaneously to its subgroup $H$, the low-energy dynamics is dominated by the ensuing NG bosons. Their scattering amplitudes and other low-energy observables can be extracted from the Green's functions of the Noether currents. Introducing a set of background gauge fields $A^i_\mu(x)$, coupled to the respective currents, the connected components of these Green's functions can be collected in a generating functional that we will denote as $\Gamma\{A\}$.\footnote{We follow the notation introduced by Leutwyler~\cite{Leutwyler:1993iq} and denote by curly brackets a nonlocal \emph{functional} of $A$. Square brackets will, on the other hand, indicate a local \emph{function} of $A$ and its derivatives.}
\begin{theorem}[Ward identities]
\label{thm:generatingfunctional}
In the absence of quantum anomalies and explicit symmetry breaking, the symmetry of the theory under the group $G$ is encoded in the invariance of the generating functional $\Gamma\{A\}$ under a gauge transformation of the background fields,
\end{theorem}
\begin{equation}
\gtr\g{A_\mu}=\g A_\mu\g^{-1}+\imag\g\de_\mu\g^{-1}.
\label{gaugetransfo_large}
\end{equation}
Here $A_\mu=A^i_\mu T_i$, and $\g\in G$ is coordinate-dependent. When $\g$ is characterized by a set of infinitesimal parameters $\epsilon^i$, $\g=e^{\imag\epsilon^iT_i}$, the transformation rule (to linear order in $\epsilon^i$) takes the more familiar form
\begin{equation}
\delta A^i_\mu=f^i_{jk}A^j_\mu\epsilon^k+\de_\mu\epsilon^i,
\label{gaugetransfo_small}
\end{equation}
where $f^i_{jk}$ are the structure constants of $G$. The low-energy observables are described equally well by an EFT which is defined by a local action, $S_\text{eff}$, in terms of the NG fields $\pi^a(x)$, one for each broken symmetry generator $T_a$. Coupling the EFT to the same background gauge fields $A^i_\mu$, it must reproduce the generating functional of the underlying microscopic theory by means of a functional integration over the NG fields,
\begin{equation}
e^{\imag\Gamma\{A\}}=\frac1Z\int\mathcal D\pi\,e^{\imag S_\text{eff}\{\pi,A\}}.
\label{genfunc}
\end{equation}
Our main task is to construct the effective action $S_\text{eff}$, or the corresponding local effective Lagrangian, given by $S_\text{eff}\{\pi,A\}=\int\dd x\,\La_\text{eff}[\pi,A]$. It is customary, especially in high energy physics, to \emph{assume} that the Lagrangian is invariant under the group $G$. However, it is far from trivial to see what the invariance of the generating functional, ensured by theorem~\ref{thm:generatingfunctional}, actually implies for the effective action $S_\text{eff}$. This problem was considered by Leutwyler, who proved the following set of statements (abbreviated; see ref.~\cite{Leutwyler:1993iq} for the full formulation), valid to all orders in the derivative expansion:
\begin{theorem}[Action invariance]
\label{thm:invariancetheorem}
(i) There exists a mapping of the NG fields, $\pi^a\xrightarrow{\g}f^a[\g,\pi,A]$ under which, together with the gauge transformation~\eqref{gaugetransfo_large} of the external fields, the action $S_\text{eff}\{\pi,A\}$ remains invariant, $S_\text{eff}\{f[\g,\pi,A],\gtr\g A\}=S_\text{eff}\{\pi,A\}$. (ii) The map $f^a[\g,\pi,A]$ defines a nonlinear realization of the group $G$, that is, obeys the composition law $f^a[\g_2\g_1,\pi,A]=f^a[\g_2,f[\g_1,\pi,A],\gtr{\g_1}A]$. (iii) With a suitable change $\pi^a\to\tilde \pi^a[\pi,A]$ of field variables, the map can be brought to certain canonical form (introduced below). In these variables, the transformation law of the NG fields is determined solely by the geometry of the group $G$ and is independent of the background fields $A^i_\mu$.
\end{theorem}
In brief, by a suitable \emph{choice} of field variables and the transformation law for NG fields, the effective action can be made invariant under a simultaneous gauge transformation of the NG and background gauge fields. Leutwyler presents his argument in the framework of relativistic field theory and asserts in addition that in four spacetime dimensions, the effective Lagrangian itself is necessarily gauge-invariant. The above-listed first three parts of his invariance theorem do not require Lorentz invariance though, and can therefore be used without modification in the more general context of quantum many-body systems.

The ``canonical'' nonlinear realization of the symmetry group, asserted by theorem~\ref{thm:invariancetheorem}, is defined as follows. Introduce an equivalence relation between two elements of $G$ under right multiplication by an element of the unbroken subgroup $H$: $\g_1$ and $\g_2$ are equivalent if and only if $\g_1=\g_2\h$ for some $\h\in H$. The set of equivalence classes with respect to this relation is called the (left) \emph{coset space} and denoted as $G/H$. Introducing the notation for the coset generated by a group element $\g$, $\cst\g=\{\g\h\,|\,\h\in H\}$, one can define a natural action of the group $G$ on the coset space $G/H$ by left multiplication, $\cst{\g'}\xrightarrow\g\cst{\g\g'}$. The subgroup $H$ forms a coset, $H=\cst\e$, which is left intact by the action of $H$ itself. In physical terms, the coset $\cst\e$ represents the vacuum, invariant by assumption under the subgroup $H$. It is convenient to pick a unique element $\uu\in\cst{}$ to represent every coset. The group action on the coset space then takes the form $\uu\xrightarrow{\g}\g\uu=\uu'\h(\uu,\g)$ where $\h(\uu,\g)\in H$ ensures that $\uu'$ coincides with the representative element of the coset $\cst{\g\uu}$. The NG fields $\pi^a$ can now be thought of as coordinates on the coset space $G/H$. Interpreting the coset element as a matrix, $\uu=U(\pi)$, the transformation law for the NG fields takes finally the usual form
\begin{equation}
U(\pi')=\g U(\pi)\h(\pi,\g)^{-1}.
\label{NGtransfo}
\end{equation}
It is common to parameterize the coset element specifically as $U(\pi)=e^{\imag\pi^aT_a}$. We would therefore like to stress that our results throughout the paper apply to fairly arbitrary parameterizations of $U(\pi)$, or fairly arbitrary choices of the NG field variables, unless explicitly stated otherwise. Namely, the only universal technical requirement is that the trivial coset $\cst\e=H$ is represented by the unit matrix and corresponds to the origin in the NG space, $U(0)=\openone$, which implies that $\h(0,\g)=\g$ for all $\g\in H$.


\subsection{General invariant actions}
\label{subsec:generalactions}

The next crucial step in the construction is the observation~\cite{Leutwyler:1993iq} that the dependence of the effective action on the NG fields and on the background gauge fields is closely related. Indeed, by choosing $\g=U(\pi)^{-1}$, we can make the NG fields vanish. Gauge invariance of the effective action, ensured by theorem~\ref{thm:invariancetheorem}, then implies
\begin{equation}
S_\text{eff}\{\pi,A\}=S_\text{eff}\{0,\gtr{U(\pi)^{-1}}A\}.
\label{gaugeinvartrick}
\end{equation}
The action is therefore fixed solely by its dependence on the gauge field. Given that the vacuum $\pi=0$ is $H$-invariant, this dependence is constrained by gauge invariance with respect to the unbroken subgroup $H$. Conversely, every $H$-invariant functional $F\{A\}$ can be used to define the effective action as $S_\text{eff}\{\pi,A\}=F\{\gtr{U(\pi)^{-1}}A\}$. That this is indeed invariant under the full group $G$ follows from eq.~\eqref{NGtransfo},
\begin{equation}
\begin{split}
S_\text{eff}\{\pi',A'\}&=F\{\gtr{U(\pi')^{-1}}{\gtr\g A}\}=F\{\gtr{U(\pi')^{-1}\g}A\}=F\{\gtr{\h(\pi,\g)U(\pi)^{-1}}A\}\\
&=F\{\gtr{\h(\pi,\g)}{\gtr{U(\pi)^{-1}}A}\}=F\{\gtr{U(\pi)^{-1}}A\}=S_\text{eff}\{\pi,A\}.
\end{split}
\label{proofgaugetrick}
\end{equation}
The problem of finding the most general $G$-invariant effective action for $\pi^a$ and $A^i_\mu$ therefore reduces to finding the most general $H$-invariant action for the gauge field alone. To that end, it is natural to split the gauge field $\gtr{U(\pi)^{-1}}A_\mu$ into components in the subspaces of the broken and unbroken generators, respectively, denoted as $\phi_\mu$ and $B_\mu$ and defined by
\begin{equation}
\gtr{U(\pi)^{-1}}A_\mu=\phi_\mu(\pi)+B_\mu(\pi)=\phi^a_\mu(\pi)T_a+B^\alpha_\mu(\pi)T_\alpha.
\label{Bphidef}
\end{equation}
From eq.~\eqref{gaugetransfo_large} we readily obtain their transformation properties under $H$,
\begin{equation}
\begin{gathered}
\gtr\h{\phi_\mu}=\h\phi_\mu\h^{-1},\qquad\gtr\h{B_\mu}=\h B_\mu\h^{-1}+\imag\h\de_\mu\h^{-1},\\
\delta\phi^a_\mu=f^a_{b\alpha}\phi^b_\mu\epsilon^\alpha,\qquad\delta B^\alpha_\mu=f^\alpha_{\beta\gamma}B^\beta_\mu\epsilon^\gamma+\de_\mu\epsilon^\alpha.
\end{gathered}
\label{Bphitransfo}
\end{equation}
This means that while $B^\alpha_\mu$ transforms as a genuine gauge field of $H$, $\phi^a_\mu$ rather behaves as a set of covariant vector fields. Altogether, using eq.~\eqref{gaugeinvartrick}, we obtain a simple algorithm for the construction of the effective action.
\begin{theorem}[Action reconstruction]
Find the most general action for the field $\phi^a_\mu$ and the auxiliary gauge field $B^\alpha_\mu$, invariant under the gauge $H$-transformations~\eqref{Bphitransfo}. The most general $G$-invariant effective action for the NG fields $\pi^a$ and the original gauge fields $A^i_\mu$ is obtained by the replacement
\end{theorem}
\begin{equation}
\begin{split}
\phi^a_\mu&\to[U(\pi)^{-1}A_\mu U(\pi)+\imag U(\pi)^{-1}\de_\mu U(\pi)]^a,\\
B^\alpha_\mu&\to[U(\pi)^{-1}A_\mu U(\pi)+\imag U(\pi)^{-1}\de_\mu U(\pi)]^\alpha.
\end{split}
\end{equation}
We have already succeeded in reformulating the problem in terms of elementary field theory, without referring to the geometry of the coset space $G/H$ and the nonlinear transformation law for the NG fields. However, the solution is still not completely straightforward. The subtlety lies in the fact that standard field theory methods allow us to construct an invariant Lagrangian, yet invariance of the action only requires that the Lagrangian be invariant up to a surface term. This is not a mere technicality: a term in the Lagrangian invariant only up to a total time derivative is responsible for the quadratic dispersion relation of some NG bosons and for their number differing from the number of broken symmetry generators~\cite{Leutwyler:1993gf,Watanabe:2012hr,Hidaka:2012ym} (see ref.~\cite{Brauner:2010wm} for a review). The problem can be further simplified by observing that the possible surface term induced by a symmetry transformation only affects a part of the Lagrangian, independent of the covariant field $\phi^a_\mu$.
\begin{theorem}[Lagrangian invariance]
\label{thm:Laginv}
The most general $H$-invariant action for the fields $\phi^a_\mu$ and $B^\alpha_\mu$ takes the form $\int\dd x\,(\La_\text{inv}[\phi,B]+\La_\text{CS}[B])$, where the Lagrangian $\La_\text{inv}[\phi,B]$ is strictly gauge-invariant under $H$.
\end{theorem}
In order to understand this statement, note that the scalar current, defined by
\begin{equation}
\Sigma^\mu_a[\phi,B]=\frac{\delta S_\text{eff}\{\phi,B\}}{\delta\phi^a_\mu},
\end{equation}
transforms covariantly under the gauge transformation~\eqref{Bphitransfo}, namely $\delta\Sigma_a^\mu=-f^b_{a\alpha}\Sigma^\mu_b\epsilon^\alpha$ (see appendix~\ref{app:covariance} for a detailed proof). Using the fact that the functional derivative of the action can be traded for an ordinary derivative with respect to a parameter,
\begin{equation}
\frac{\de S_\text{eff}\{t\phi,B\}}{\de t}=\int\dd x\,\phi^a_\mu\Sigma_a^\mu[t\phi,B],
\end{equation}
we can reconstruct the action by an integration of the current over this parameter,
\begin{equation}
S_\text{eff}\{\phi,B\}=S_\text{eff}\{0,B\}+\int\dd x\int_0^1\dd t\,\phi^a_\mu\Sigma^\mu_a[t\phi,B].
\label{Sinv}
\end{equation}
The transformation rule for $\phi^a_\mu$ is homogeneous, hence the argument of the coordinate integral above defines a gauge-invariant Lagrangian density, $\La_\text{inv}[\phi,B]$. The remaining part of the action depends solely on the gauge field $B^\alpha_\mu$, as we wanted to show.

Let us now focus on the term $\La_\text{CS}[B]$, which represents a gauge theory whose Lagrangian may change under the gauge transformation~\eqref{Bphitransfo} by a surface term, and therefore constitutes a generalization of the Chern-Simons theory. We will henceforth refer to such terms in the Lagrangian as Chern-Simons (CS). The construction of the CS part of the Lagrangian follows the same steps as sketched above in the case of $\La_\text{inv}[\phi,B]$. Namely, the current
\begin{equation}
J^\mu_\alpha[B]=\frac{\delta S_\text{CS}\{B\}}{\delta B^\alpha_\mu}
\end{equation}
again transforms covariantly, that is, $\delta J^\mu_\alpha=-f^\gamma_{\alpha\beta}J^\mu_\gamma\epsilon^\beta$ (see appendix~\ref{app:covariance} for a proof), and the Lagrangian can subsequently be reconstructed as
\begin{equation}
\La_\text{CS}[B]=\int_0^1\dd t\,B^\alpha_\mu J^\mu_\alpha[tB].
\label{CSmaster}
\end{equation}
The essential difference to eq.~\eqref{Sinv} is that the gauge field does not transform homogeneously and thus the Lagrangian density is now not necessarily gauge-invariant.


\subsection{Construction of effective Lagrangians}
\label{subsec:Lagrangianconstruction}

We can conclude that the construction of the effective theory reduces to the classification of certain gauge-covariant objects: the Lagrangian densities in case of $\La_\text{inv}[\phi,B]$ and the currents in case of $\La_\text{CS}[B]$. These can be obtained using common field-theoretical methods. To be precise, let us denote as \emph{gauge-covariant} a local function of the fields $\phi^a_\mu,B^\alpha_\mu$ and their derivatives, whose infinitesimal shift under the gauge transformation~\eqref{Bphitransfo} does not contain derivatives of the parameters $\epsilon^\alpha$. We use the following well-known statement.
\begin{theorem}[Covariance of building blocks]
\label{thm:covariantobjects}
Consider a set of gauge fields $A^i_\mu$, and of matter fields $\phi^a$ transforming in a given linear representation $R$ of the gauge group. Every local gauge-covariant function of $\phi^a$ and $A^i_\mu$ and their derivatives can be expressed solely in terms of $\phi^a$, its covariant derivative $D_\mu\phi^a=\de_\mu\phi^a-\imag R(A_\mu)^a_{\phantom ab}\phi^b$, the field-strength tensor $F^i_{\mu\nu}=\de_\mu A^i_\nu-\de_\nu A^i_\mu+f^i_{jk}A^j_\mu A^k_\nu$, and their covariant derivatives.
\end{theorem}
This is a standard textbook result, yet it does not seem easy to find its proof in the full generality required here in the literature. For the sake of completeness and for the reader's convenience, we provide a detailed argument in appendix~\ref{app:theoremproof}.

The construction of both the invariant Lagrangian $\La_\text{inv}[\phi,B]$ and the covariant current $J^\mu_\alpha[B]$ now proceeds as follows. First, we find all linearly independent operators, $\mathcal O_A$, as products of the basic building blocks ($\phi^a_\mu$, $G^\alpha_{\mu\nu}$ and their covariant derivatives) that contribute at a given order of the derivative expansion; since each of the building blocks contains at least one spacetime index, there is always a finite number of such operators. The desired covariant object (Lagrangian or current) is then written as a linear combination, $\sum_Ac_A\mathcal O_A$, with unknown effective couplings $c_A$. The linear independence of the set of operators $\mathcal O_A$ guarantees that each term in the sum has to be covariant separately from the others, while the covariance of our building blocks in turn implies that the couplings, with all the group indices restored, have to be \emph{invariant tensors} of the unbroken subgroup $H$. Since (continuous) spacetime symmetries are by assumption not spontaneously broken, the building blocks also transform covariantly under those, and the effective couplings $c_A$ have to be simultaneously invariant tensors of the spacetime symmetry group.

Altogether, the classification of effective Lagrangians boils down to the enumeration of all possible operators expressed using our basic building blocks, and to elementary group theory, namely to finding all invariant tensors of $H$ and the spacetime symmetry with the appropriate number of indices of each type: $a$ coming from $\phi$, $\alpha$ coming from $B$, and $\mu$ from both. Mathematically, this amounts to taking the direct product of representations corresponding to all the fields in a given operator $\mathcal O_A$ and finding all singlets in its decomposition into irreducible components. To that end, we will often use the fact that given the invariant tensors of two groups $G_1$ and $G_2$, the invariant tensors of their product $G_1\times G_2$ can be obtained by taking all possible products of invariant tensors of the two subgroups.

In case of the invariant Lagrangians, the resulting list of possible terms can be further reduced. Since every operator $\mathcal O_A$ is separately gauge invariant, we can rewrite it using the integration by parts formula,
\begin{equation}
\int\dd x\,\mathcal O_1(D_\mu\mathcal O_2)=-\int\dd x\,(D_\mu\mathcal O_1)\mathcal O_2+\text{surface term}.
\end{equation}
Note that gauge invariance is essential to ensure that $D_\mu(\mathcal O_1\mathcal O_2)=\de_\mu(\mathcal O_1\mathcal O_2)$ here is a mere surface term. For topologically trivial field configurations, the surface term can be discarded, and we will always do so since we are primarily interested in the low-energy physics of the NG bosons.


\section{Leading-order effective Lagrangian}
\label{sec:LO}

In this section, we show in detail how the strategy outlined above can be used to work out the most general effective Lagrangian up to the second order in the derivative expansion. Owing to the simplicity of this problem, we are able to work out the solution without making \emph{any} assumptions on the spacetime symmetry.

Let us first focus on the invariant part of the Lagrangian, $\La_\text{inv}[\phi,B]$. Up to second order in derivatives, the following operators are available,
\begin{equation}
\phi^a_\mu\quad\text{(order 1)},\qquad
\phi^a_\mu\phi^b_\nu,\xcancel{D_\mu\phi^a_\nu},\xcancel{G^\alpha_{\mu\nu}}\quad
\text{(order 2)},
\label{LOoperators}
\end{equation}
where the crossed out operators do not contribute. Of them, $D_\mu\phi^a_\nu$ is a total derivative and thus constitutes just a surface term, while the reason why $G^\alpha_{\mu\nu}$ does not contribute to the action either will be explained shortly. The most general invariant Lagrangian up to second order in derivatives therefore reads
\begin{equation}
\La_\text{inv}=e^\mu_a\phi^a_\mu+\frac12g^{\mu\nu}_{ab}\phi^a_\mu\phi^b_\nu,
\end{equation}
where $e^\mu_a$ and $g^{\mu\nu}_{ab}$ are invariant tensors of the unbroken subgroup $H$, whose action on the fields is defined by eq.~\eqref{Bphitransfo}. Hence, they have to satisfy the conditions $e^\mu_af^a_{\alpha b}=0$ and $g^{\mu\nu}_{cb}f^c_{\alpha a}+g^{\mu\nu}_{ac}f^c_{\alpha b}=0$ for all allowed values of the indices, see eq.~\eqref{invariancecondition}. Likewise, they are invariant tensors of the assumed spacetime symmetry.

In order to determine the CS part of the Lagrangian up to the second order in derivatives, we need to list all possible covariant currents $J^\mu_\alpha[B]$ up to order one. Since the simplest covariant operator one can construct out of $B^\alpha_\mu$ is the field-strength tensor $G^\alpha_{\mu\nu}$ which is of order two, there is obviously only one possibility, namely a constant current $J^\mu_\alpha=e^\mu_\alpha$. The integration indicated in eq.~\eqref{CSmaster} is in this case trivial, leading to
\begin{equation}
\La_\text{CS}=e^\mu_\alpha B^\alpha_\mu,
\end{equation}
where the coupling is again an invariant tensor of $H$, that is, $e^\mu_\alpha f^\alpha_{\beta\gamma}=0$. It is now clear why the operator $G^\alpha_{\mu\nu}$ cannot contribute to the Lagrangian. Including the appropriate effective coupling, it would produce $c^{\mu\nu}_\alpha G^\alpha_{\mu\nu}$, however the $\de_\mu B^\alpha_\nu-\de_\nu B^\alpha_\mu$ part of the field-strength tensor would drop immediately being a surface term, while the non-Abelian part $c^{\mu\nu}_\alpha f^\alpha_{\beta\gamma}B^\beta_\mu B^\gamma_\nu$ would vanish by means of the invariance condition on the coupling $c^{\mu\nu}_\alpha$.

Altogether, the general leading-order effective Lagrangian together with the corresponding constraints on the effective couplings can be written as
\begin{equation}
\begin{gathered}
\La^\text{LO}_\text{eff}=e^\mu_\alpha B^\alpha_\mu+e^\mu_a\phi^a_\mu+\frac12g^{\mu\nu}_{ab}\phi^a_\mu\phi^b_\nu,\\
e^\mu_i f^i_{\alpha j}=0,\qquad g^{\mu\nu}_{cb}f^c_{\alpha a}+g^{\mu\nu}_{ac}f^c_{\alpha b}=0,
\end{gathered}
\label{LOlagrangian}
\end{equation}
where the metric $g^{\mu\nu}_{ab}$ can in addition be assumed symmetric, $g^{\mu\nu}_{ab}=g^{\nu\mu}_{ba}$. This agrees with the result obtained recently in ref.~\cite{Watanabe:2014fva}. In that paper, a generalization of the EFT to the cases where the global symmetry under the group $G$ cannot be gauged is studied. It is obvious that \emph{assuming} gauge invariance dramatically simplifies the derivation of the most general effective Lagrangian, reducing a rather elaborate calculation to a back-of-the-envelope argument.

In rotationally invariant systems, $e^\mu_i=e_i\delta^{\mu0}$. In fact, even a discrete space symmetry is sufficient to ensure this relation. In the following, we will always implicitly assume it, since a term linear in spatial derivatives would otherwise necessarily lead to a spontaneous breakdown of continuous translational invariance~\cite{Watanabe:2014fva}. Under rotational invariance, the bilinear part of the Lagrangian further reduces to $g^{\mu\nu}_{ab}\phi^a_\mu\phi^b_\nu=\bar g_{ab}\phi^a_0\phi^b_0-g_{ab}\phi^a_r\phi^b_r$. In two spatial dimensions, an additional, antisymmetric bilinear term is allowed, $\bar{\bar g}_{ab}\epsilon^{rs}\phi^a_r\phi^b_s$.


\subsection{Physical implications}
\label{subsec:LOphysical}

With the help of eq.~\eqref{auxfields}, we can re-express the Lagrangian~\eqref{LOlagrangian} in terms of the physical NG fields and the background gauge fields,
\begin{equation}
\La^\text{LO}_\text{eff}=-e^\mu_i\omega^i_a(\pi)\de_\mu\pi^a+e^\mu_j\nu^j_i(\pi)A^i_\mu+\tfrac12g^{\mu\nu}_{ab}\omega^a_c(\pi)\omega^b_d(\pi)D_\mu\pi^c D_\nu\pi^d.
\label{LOlagrangianexplicit}
\end{equation}
This Lagrangian takes the form first obtained by Leutwyler~\cite{Leutwyler:1993gf}, and features explicit expressions for his coupling functions in terms of the objects $\omega^i_a$ and $\nu^i_j$, defined in eq.~\eqref{MCform}. For the specific parameterization $U(\pi)=e^{\imag\pi^aT_a}$, we can moreover use eq.~\eqref{expparam} to obtain the expansion of the Lagrangian in powers of the 
NG fields,
\begin{equation}
\La_\text{eff}^\text{LO}=\tfrac12e_i f^i_{ab}\de_0\pi^a\pi^b+e_iA^i_0+\tfrac12g^{\mu\nu}_{ab}D_\mu\pi^aD_\nu\pi^b+\dotsb.
\label{LOlagrangianexpansion}
\end{equation}
Note that the same coupling $e_i$ appears both in the term linear in $A^i_\mu$ and the term quadratic in $\pi^a$ with a single time derivative. The former implies that $e_i$ has the meaning of the vacuum expectation value of the charge density associated with the generator $T_i$, while the latter indicates that whenever the commutator $[T_a,T_b]$ has a nonzero vacuum expectation value, the field variables $\pi^a$ and $\pi^b$ are canonically conjugated~\cite{Nambu:2004yia}. Such a pair of field variables excites \emph{one} NG boson, classified as type B~\cite{Watanabe:2012hr}; owing to the presence of a term with a single time derivative, their dispersion relation is typically quadratic in momentum~\cite{Nielsen:1975hm}. On the other hand, the remaining NG fields excite one type-A NG boson each, whose dispersion relations are, as a rule, linear in momentum.

The effective coupling in the bilinear part of the Lagrangian has a particularly simple interpretation in the rotationally invariant case. Namely, the couplings $g_{ab}$ and $\bar g_{ab}$ encode the amplitude for the creation of the NG boson by the associated broken current, usually dubbed the NG boson decay constant. Their ratio in turn determines the phase velocity of type-A NG bosons. As follows from the discussion in appendix~\ref{app:tensor}, there is one parameter of each type for every irreducible multiplet of NG bosons.

Apart from the dispersion relations of the NG bosons, the nonlinear dependence of the Lagrangian~\eqref{LOlagrangianexplicit} on $\pi^a$ determines the dominant interactions of NG bosons at low energy or momentum. We emphasize what should already be clear from the above equations: this nonlinear dependence is fixed by symmetry, and the low-energy physics of NG bosons is fully determined by the set of leading-order effective couplings, that is, their decay constants and phase velocities, and the charge densities in the ground state.


\subsection{Power counting}
\label{subsec:powercounting}

Now that we have discussed the spectrum of NG bosons, we return to the question of power counting, which determines how the derivative expansion of the Lagrangian is organized. So far, we sorted the Lagrangian separately by the number of spatial and temporal indices. However, for a well-defined expansion, one needs a unique expansion parameter. In fact,  we have so far been discussing the leading-order Lagrangian without having a clear notion of what ``leading-order'' means.

Let us recall how the powers of derivatives are counted in $\chi$PT~\cite{Scherer:2002tk}, or in general Lorentz-invariant systems. There, spatial and temporal derivatives are treated on the same footing, and each of them, as well as the background fields $A^i_\mu$, is counted as order one. The propagator of a NG boson is then of order $-2$ and consequently a given Feynman diagram with $L$ loops and $I$ propagators in $d$ spacetime dimensions has the superficial degree of divergence~\cite{Burgess:1998ku}
\begin{equation}
\deg=dL-2I+\sum_vd_v=2+(d-2)L+\sum_v(d_v-2),
\label{powercountingA}
\end{equation}
where $d_v$ denotes the number of derivatives in the operator representing the vertex $v$. Since every operator in the Lagrangian contains at least two derivatives, the power counting is well defined. The leading contribution to any Green's function or scattering amplitude is of order two, and constitutes solely tree-level diagrams with vertices from $\La^{(2)}_\text{eff}$. Adding loops (in $d\geq3$) or vertices from higher-order operators increases the order. This guarantees that, to any finite order in the derivative expansion, only a finite number of operators and Feynman diagrams contribute. In the exceptional case of $d=2$, adding loops does not increase the order of the diagram. This is another manifestation of the strong infrared fluctuations of the NG fields which eventually lead to the restoration of the symmetry~\cite{Coleman:1973ci}.

The above argument applies to all systems with a purely type-A NG boson spectrum. The fact that the phase velocities of the NG bosons are not equal to the speed of light does not need to concern us: all that matters is that the energy scales linearly with momentum. As a consequence, the part of the effective Lagrangian $\La^{(s,t)}_\text{eff}$ is assigned the order $s+t$, which enters eq.~\eqref{powercountingA} through the vertex degree $d_v$.

Let us now consider the opposite extreme, namely a system in which all NG bosons are of type B and have a quadratic dispersion relation, such as~a ferromagnet. In order for the NG boson propagator to have a well-defined degree, each temporal derivative now has to be counted as two spatial derivatives. It is therefore natural to count $\de_r$ as order one, and $\de_0$ as order two. The power-counting formula~\eqref{powercountingA} then changes accordingly,
\begin{equation}
\deg=(d+1)L-2I+\sum_vd_v=2+(d-1)L+\sum_v(d_v-2),
\label{powercountingB}
\end{equation}
where $d_v$ is now the total order of the vertex $v$, taking into account the difference between spatial and temporal indices. The same argument asserting the existence of a well-defined power counting as above applies, except that now one has a valid derivative expansion even at $d=2$~\cite{Hofmann:2012uy}. The Lagrangian $\La^{(s,t)}_\text{eff}$ is correspondingly assigned the total order $s+2t$.

The above lengthy considerations finally define the notion of a leading order in our expansion. Barring the occurrence of operators with a single spatial derivative, this always carries two powers of momentum. In pure type-A systems, the leading-order Lagrangian contains terms with two spatial or two temporal derivatives. In pure type-B systems such as ferromagnets, it contains terms with two spatial or \emph{one} time derivative. Operators with two temporal derivatives, implicit in eq.~\eqref{LOlagrangianexplicit}, are then only subleading, of order four.

Our discussion suggests a natural question: how to define power counting in mixed systems where both types of NG bosons appear? This is not merely an academic question; such systems include for instance the canted phase of ferromagnets~\cite{Roman:canted} or certain models of relativistic Bose-Einstein condensation~\cite{Miransky:2001tw,Schafer:2001bq}. For operators built solely out of $\phi^a_\mu$ and $G^\alpha_{\mu\nu}$ without additional derivatives, one can alternatively assign a fixed order directly to the respective component of the MC form, $\phi^a_\mu$ or $B^\alpha_\mu$. If the field $\pi^a$ belongs to a pair of variables canonically conjugated by the coupling $e_i$, $\phi^a_0$ is counted as order two, otherwise it is assigned the order one. However, this still does not give a unique prescription for operators carrying extra covariant derivatives $D_\mu$, since these can be moved by partial integration within a product of fields. How to define power counting in this general case remains a problem to be resolved in the future.


\subsection{Equation of motion}
\label{subsec:EOM}

It is instructive to find the equation of motion stemming from the leading-order Lagrangian~\eqref{LOlagrangian}. As we argue below, this allows one to eliminate some of the numerous operators contributing at higher orders of the derivative expansion. Also, it may help to elucidate the spectrum of the NG modes and other physical observables such as the response of the system to external fields. In this section, we sketch its derivation in a form manifestly covariant under all the symmetries.

It is convenient to collect the auxiliary fields $\phi^a_\mu$ and $B^\alpha_\mu$ in a single variable, $\tilde A_\mu=\tilde A^i_\mu T_i=\gtr{U^{-1}}A_\mu=\phi_\mu+B_\mu$. In terms of the objects introduced in eq.~\eqref{MCform}, it reads
\begin{equation}
\tilde A_\mu(\pi)=U(\pi)^{-1}(A_\mu+\imag\de_\mu)U(\pi)=[A^j_\mu\nu^i_j(\pi)-\omega^i_a(\pi)\de_\mu\pi^a]T_i.
\end{equation}
We know from eq.~\eqref{gaugeinvartrick} that the effective action can be expressed solely in terms of $\tilde A_\mu(\pi)$. The general equation of motion is then
\begin{equation}
0=\frac{\delta S_\text{eff}\{\pi,A\}}{\delta\pi^a(x)}=\int\dd y\,\frac{\delta S_\text{eff}}{\delta\tilde A^i_\mu(y)}\frac{\delta\tilde A^i_\mu(y)}{\delta\pi^a(x)}.
\end{equation}
Using the above expression for $\tilde A^i_\mu$ in terms of $\omega^i_a$ and $\nu^i_j$, it is straightforward to evaluate the second of the functional derivatives under the integral. In order to bring the equation of motion into a covariant form, one can in addition use the identity $\de_a\nu_i=-\imag[\omega_a,\nu_i]$ and the Maurer-Cartan structure equation~\cite{Nakahara},
\begin{equation}
\de_a\omega_b-\de_b\omega_a=-\imag[\omega_a,\omega_b],\qquad\text{or}\quad\de_a\omega^i_b-\de_b\omega^i_a=f^i_{jk}\omega^j_a\omega^k_b,
\label{MCequation}
\end{equation}
both of which are easily obtained from the definitions of $\omega^i_a$ and $\nu^i_j$ in eq.~\eqref{MCform}. The equation of motion then acquires the form
\begin{equation}
\omega^j_a\left(\delta^i_j\de_\mu\frac{\delta S_\text{eff}}{\delta\tilde A^i_\mu}+f^i_{jk}\tilde A^k_\mu\frac{\delta S_\text{eff}}{\delta\tilde A^i_\mu}\right)=0.
\label{EOMgeneral}
\end{equation}

Let us now see what this implies at the lowest orders of the derivative expansion. In terms of $\tilde A^i_\mu$ the Lagrangian~\eqref{LOlagrangian} takes the rather compact form $\La^\text{LO}_\text{eff}=e^\mu_i\tilde A^i_\mu+\frac12g^{\mu\nu}_{ab}\tilde A^a_\mu\tilde A^b_\nu$. Plugging this into eq.~\eqref{EOMgeneral} and expanding in all possible combinations of broken and unbroken indices seems to produce a lot of terms. However, some of them vanish due to the invariance conditions: (i) $\omega^j_af^i_{jk}\tilde A^k_\mu e^\mu_i$ vanishes unless both $j$ and $k$ are broken indices; (ii) $\omega^\alpha_af^b_{\alpha k}\tilde A^k_\mu(g^{\mu\nu}_{bc}\tilde A^c_\nu)=\omega^\alpha_af^b_{\alpha d}g^{\mu\nu}_{bc}\tilde A^d_\mu\tilde A^c_\nu$ vanishes since as follows from appendix~\ref{subapp:cab}, $f^b_{\alpha d}g^{\mu\nu}_{bc}$ is antisymmetric under the simultaneous exchange of $c,d$ and $\mu,\nu$. Upon some relabeling of the indices, the equation of motion then reduces to
\begin{equation}
\omega^b_a(g^{\mu\nu}_{bc}\de_\mu\tilde A^c_\nu+f^c_{bi}g^{\mu\nu}_{cd}\tilde A^i_\mu\tilde A^d_\nu+e^\mu_if^i_{bc}\tilde A^c_\mu)=0.
\end{equation}
From eq.~\eqref{expparam} we know that at $\pi=0$, $\omega^b_a(0)=\delta^b_a$. Therefore, there is a neighborhood of the origin of the coset space $G/H$ in which the matrix $\omega^b_a(\pi)$ is nondegenerate.\footnote{This conclusion is independent of the specific choice of parameterization, used to derive eq.~\eqref{expparam}.} In this neighborhood, we  can divide the entire equation by this factor. Upon finally splitting the index $i$ in the second term into its broken and unbroken part, we realize that the equation of motion can be cast in a manifestly $G$-invariant form, solely in terms of $\phi^a_\mu$,
\begin{equation}
f^i_{ab}e^\mu_i\phi^b_\mu+g^{\mu\nu}_{ab}D_\mu\phi^b_\nu+f^b_{ac}g^{\mu\nu}_{bd}\phi^c_\mu\phi^d_\nu=0.
\label{EOMLO}
\end{equation}
This form of the equation of motion does not make any assumptions on the spacetime symmetry, and constitutes a generalization of the Landau-Lifschitz equation for the spin waves in ferromagnets~\cite{Leutwyler:1993gf}. Of course, in most cases of physical interest, it takes a particularly simple form. First, in rotationally invariant systems (in two or more spatial dimensions), the first term reduces to $f^i_{ab}e_i\phi^b_0$. Second, in rotationally invariant systems in three spatial dimensions, the second term reduces to $\bar g_{ab}D_0\phi^b_0-g_{ab}D_r\phi^b_r$, that is, to the covariant Laplacian. Finally, the last term is missing when the coset space $G/H$ is symmetric.

The equation of motion~\eqref{EOMLO} is expressed in terms of the same building blocks as the invariant part of the Lagrangian $\La_\text{inv}$, allowing us to eliminate some of the operators that appear in the higher orders of the derivative expansion. This is equivalent to a certain field redefinition, and therefore provides a tool to reduce redundancy in the higher-order Lagrangians~\cite{Fearing:1994ga,Bijnens:1999sh}. How precisely this procedure works again depends on the classification of the NG bosons. For a type-B NG boson $\pi^a$, $\phi^a_0$ appears linearly in eq.~\eqref{EOMLO}, hence it can be eliminated altogether from the higher-order operators. For all the remaining generators, corresponding to type-A NG bosons, $D_0\phi^a_0$ can be eliminated in favor of $D_r\phi^a_r$ and products of $\phi^a_\mu$s. In the special case of Lorentz-invariant systems, the equation of motion allows one to remove operators containing $D_\mu\phi^{a\mu}$.


\section{Higher-order contributions}
\label{sec:higherorders}

In this section, we provide some details of the construction of the order-four effective Lagrangian presented in section~\ref{sec:summary}. It is worthwhile to stress the conceptual simplicity of our approach: we merely have to list all possible operators up to the desired order and find all invariant tensors of the unbroken subgroup $H$ to contract their indices. The nontrivial task turns out not to be to make sure that the list of operators is complete, but to detect possible redundancies. There are several tools that allow one to relate apparently different operators for general $G$ and $H$~\cite{Fearing:1994ga,Bijnens:1999sh} : integration by parts, Bianchi identity for the field-strength tensor $G^\alpha_{\mu\nu}$, Schouten identity for the fully antisymmetric tensor $\epsilon^{\lambda\mu\nu\dotsb}$, and the use of the leading-order equation of motion~\eqref{EOMLO}. In addition, specific algebraic relations of the symmetry group or its representations may give rise to further constraints.


\subsection{Invariant terms}
\label{subsec:invariant}

In order to obtain explicit expressions for the invariant part of the Lagrangian, we restrict ourselves to rotationally invariant theories. However, if needed, our approach can be straightforwardly applied to more complicated cases such as condensed matter systems with a discrete space group. The problem to solve is then as easy as it gets: find the appropriate invariant tensors of the unbroken subgroup $H$.

Following the steps outlined in section~\ref{sec:method} and applied to the lowest orders of the derivative expansion in section~\ref{sec:LO}, we first list all types of operators that contribute at orders three and four:
\begin{equation}
\begin{split}
\text{order 3: }&\phi^a_\lambda\phi^b_\mu\phi^c_\nu,\phi^a_\lambda D_\mu\phi^b_\nu,\xcancel{D_\lambda D_\mu\phi^a_\nu},\phi^a_\lambda G^\alpha_{\mu\nu},\xcancel{D_\lambda G^\alpha_{\mu\nu}}.\\
\text{order 4: }&\phi^a_\kappa\phi^b_\lambda\phi^c_\mu\phi^d_\nu,\phi^a_\kappa\phi^b_\lambda D_\mu\phi^c_\nu,\xcancel{\phi^a_\kappa D_\lambda D_\mu\phi^b_\nu},D_\kappa\phi^a_\lambda D_\mu\phi^b_\nu,\xcancel{D_\kappa D_\lambda D_\mu\phi^a_\nu},\\
&\phi^a_\kappa\phi^b_\lambda G^\alpha_{\mu\nu},D_\kappa\phi^a_\lambda G^\alpha_{\mu\nu},\xcancel{\phi^a_\kappa D_\lambda G^\alpha_{\mu\nu}},\xcancel{D_\kappa D_\lambda G^\alpha_{\mu\nu}},G^\alpha_{\kappa\lambda}G^\beta_{\mu\nu}.
\end{split}
\label{listofoperators}
\end{equation}
The crossed out operators are redundant since they are either total derivatives, or can be eliminated in favor of the remaining operators using integration by parts. As the next step, we have to determine all contractions of the Lorentz indices that are allowed by the assumed spacetime symmetry. Here, it is convenient to discuss separately the case of one spatial dimension, where Lorentz invariance is not an issue for spontaneous symmetry breaking does not occur in one-dimensional Lorentz-invariant systems~\cite{Coleman:1973ci}. Since there is no (continuous) spacetime symmetry in this case, the most straightforward approach is to simply assign the temporal and spatial indices $0,1$ to the above operators in all possible ways. Along the way, one encounters further redundancies as some of the operators can be eliminated using integration by parts. For instance, $\phi^a_0\phi^b_0D_0\phi^c_1$ does not appear among the operators contributing to $\La^{(1,3)}_\text{inv}$, shown in section~\ref{subsubsec:Linvar}, as it can be integrated by parts to $\phi^a_0\phi^b_1D_0\phi^c_0$ (but not vice versa). Similar reasoning allows us to eliminate other operators.

In higher dimensions, one can follow two approaches, resulting in somewhat different classifications of operators in the Lagrangian. The first approach relies on the fact that the only algebraically independent invariant tensors of the Lorentz group are the Minkowski metric $\eta_{\mu\nu}$ and the Levi-Civita tensor $\epsilon_{\lambda\mu\nu\dotsb}$. The assumed rotational invariance is taken into account by introducing an additional invariant: a time-like rest-frame vector, $n^\mu=(1,0,0,\dotsc)$. One next has to enumerate all tensors obtained by products of $\eta_{\mu\nu},\epsilon_{\lambda\mu\nu\dotsb},n_\mu$ containing the desired total number of indices. Bearing in mind that a product of two $\epsilon$'s can be decomposed into a linear combination of products of the $\eta$'s, the full list of rotationally invariant tensors up to order four reads:
\begin{align}
\notag
\text{order 1: }&n_\mu.\\
\label{listoftensors}
\text{order 2: }&\underline{\eta_{\mu\nu}}, n_\mu n_\nu, \twod{\epsilon_{\mu\nu\sigma}n^\sigma}.\\
\notag
\text{order 3: }&\eta_{\lambda\mu}n_\nu, n_\lambda n_\mu n_\nu, \underline{\twod{\epsilon_{\lambda\mu\nu}}}, \twod{n_\lambda\epsilon_{\mu\nu\sigma}n^\sigma}, \threed{\epsilon_{\lambda\mu\nu\sigma}n^\sigma}.\\
\notag
\text{order 4: }&\underline{\eta_{\kappa\lambda}\eta_{\mu\nu}}, \eta_{\kappa\lambda}n_\mu n_\nu, n_\kappa n_\lambda n_\mu n_\nu , \twod{\epsilon_{\kappa\lambda\mu}n_\nu}, \twod{\epsilon_{\kappa\lambda\sigma}n^\sigma \eta_{\mu\nu}}, \twod{\epsilon_{\kappa\lambda\sigma}n^\sigma n_\mu n_\nu}, \underline{\threed{\epsilon_{\kappa\lambda\mu\nu}}}, \threed{n_\kappa\epsilon_{\lambda\mu\nu\sigma}n^\sigma}.
\end{align}
As before, we use color coding to highlight tensors that are only available for certain spatial dimensionality; moreover, tensors that are explicitly Lorentz-invariant are highlighted by underlining. 

While the above is certainly a \emph{complete} list of rotationally invariant tensors, it is, unfortunately, not \emph{minimal}. To understand why, take the antisymmetric tensor $\epsilon_{\mu_1\dotsb\mu_d}$ in $d$ spacetime dimensions, and construct a tensor of rank $d+1$ as $\epsilon_{\mu_1\dotsb\mu_d}n_{\mu_{d+1}}$. By summing over cyclic permutations of the indices, we obtain a rank-$(d+1)$ antisymmetric tensor which must identically vanish,
\begin{equation}
\epsilon_{\mu_1\dotsb\mu_d}n_{\mu_{d+1}}+(-1)^d\epsilon_{\mu_2\dotsb\mu_d\mu_{d+1}}n_{\mu_1}+\epsilon_{\mu_3\dotsb\mu_{d+1}\mu_1}n_{\mu_2}+\dotsb+(-1)^d\epsilon_{\mu_{d+1}\mu_1\dotsb\mu_{d-1}}n_{\mu_d}=0.
\end{equation}
This is a particular example of the Schouten identity. Multiplying the whole equation by $n^{\mu_{d+1}}$, we then obtain a constraint relating tensors of rank $d$, listed above,
\begin{equation}
(-1)^{d+1}\epsilon_{\mu_1\dotsb\mu_d}=\sum_{k=1}^d(-1)^{(d+1)(k+1)}n_{\mu_k}\epsilon_{\mu_{k+1}\dotsb\mu_d\mu_1\dotsb\mu_{k-1}\nu}n^\nu.
\end{equation}
This means that a manifestly Lorentz-invariant tensor $\epsilon_{\mu_1\dotsb\mu_d}$ can be recovered from a linear combination of other, Lorentz-noninvariant tensors. We have in principle two options: either keep manifest Lorentz invariance, or keep the set of tensors minimal by dropping $\epsilon_{\lambda\mu\nu}$ and $\epsilon_{\kappa\lambda\mu\nu}$ from the above list. In view of the applications of the formalism in condensed matter physics, we use the former approach to work out the effective Lagrangian only in Lorentz-invariant theories, see section~\ref{subsubsec:LLI}. There, the result is obtained by merely contracting all of the operators in eq.~\eqref{listofoperators} with the tensors of eq.~\eqref{listoftensors} at the given order.

In nonrelativistic systems one has to treat spatial and temporal indices separately, and using the Lorentz-covariant formalism proposed above would only obscure the power counting. We therefore use a different strategy. Instead of Lorentz-covariant tensors, we list all rotationally invariant tensors with spatial indices only:
\begin{equation}
\begin{split}
\text{order 2: }&\delta_{rs},\twod{\epsilon_{rs}}.\\
\text{order 3: }&\threed{\epsilon_{rst}}.\\
\text{order 4: }&\delta_{rs}\delta_{tu},\twod{\delta_{rs}\epsilon_{tu}}.
\end{split}
\end{equation}
These are again put together with operators from eq.~\eqref{listofoperators} in all possible ways, except that not all indices now have to be contracted; the leftover ones are assigned the value $0$. This way, one obtains the list of operators presented in section~\ref{subsubsec:Linvar}.

In the process, we again encounter a number of additional redundancies that somewhat reduce the final list of operators. Let us point out some of them explicitly:
\begin{itemize}
\item Some of the operators can obviously be expressed in terms of others using integration by parts. For instance, $\phi^a_r\phi^b_rD_s\phi^c_s$ can be converted into $\phi^a_r\phi^b_sD_r\phi^c_s$ and thus is not independent.
\item In two spatial dimensions, $G^\alpha_{rs}=G^\alpha_{12}\epsilon_{rs}$, therefore the naively anticipated operators $\epsilon^{st}\phi^a_r\phi^b_sG^\alpha_{rt}$ and $\epsilon^{st}D_r\phi^a_sG^\alpha_{rt}$ reduce to the operators $\epsilon^{st}\phi^a_r\phi^b_rG^\alpha_{st}$ and $\epsilon^{st}D_r\phi^a_rG^\alpha_{st}$.
\item In three spatial dimensions, the operator $\epsilon^{rst}D_r\phi^a_0G^\alpha_{st}$ vanishes through the Bianchi identity when integrated by parts.
\item The two-dimensional operator $\epsilon^{rs}D_r\phi^a_0G^\alpha_{0s}$ can be expressed in terms of $\epsilon^{rs}D_0\phi^a_0G^\alpha_{rs}$, and the three-dimensional operator $\epsilon^{rst}D_r\phi^a_sG^\alpha_{0t}$ in terms of $\epsilon^{rst}D_0\phi^a_rG^\alpha_{st}$, via the Bianchi identity. 
\item The operator $\epsilon^{rst}G^\alpha_{0r}G^\beta_{st}$ is a topological density and thus does not contribute to the action.
\end{itemize}


\subsection{Chern-Simons terms}
\label{subsec:chernsimons}

In order to determine the CS-type terms in the Lagrangian, we have to find all admissible currents $J^\mu_\alpha[B]$ and integrate them by means of eq.~\eqref{CSmaster}. The current is constrained by the requirement of covariance~\eqref{currenttransfo}, which however follows already from global symmetry alone. The assumed gauge invariance imposes an additional constraint on $J^\mu_\alpha$. To see this, simply express gauge invariance using the transformation rule~\eqref{Bphitransfo} as
\begin{equation}
0=\delta S_\text{CS}\{B\}=\int\dd x\,(f^\alpha_{\beta\gamma}B^\beta_\mu\epsilon^\gamma+\de_\mu\epsilon^\alpha)J^\mu_\alpha[B],
\end{equation}
from where an integration by parts leads us to the conservation condition
\begin{equation}
\de_\mu J^\mu_\alpha+f^\gamma_{\alpha\beta}B^\beta_\mu J^\mu_\gamma=D_\mu J^\mu_\alpha=0.
\label{currentconservation}
\end{equation}
Due to theorem~\ref{thm:covariantobjects}, the current can be constructed out of covariant building blocks. Since $S_\text{CS}\{B\}$ depends solely on the gauge field $B^\alpha_\mu$, we have $G^\alpha_{\mu\nu}$ and its covariant derivatives at our disposal. The validity of eq.~\eqref{currenttransfo} is then ensured by contracting indices in the operator with a coupling which is an invariant tensor of the unbroken subgroup $H$, leaving free one overall Lorentz index and one adjoint group index. This part of the construction is accomplished using exactly the same steps as in the case of the invariant Lagrangian $\La_\text{inv}[\phi,B]$. Finally, in order to get a gauge-invariant action, we impose the conservation condition~\eqref{currentconservation} which further restricts the possible values of the couplings.

Let us now proceed to the construction. It is worth stressing that up to order four in derivatives, the CS contributions to the Lagrangian can be worked out without making any assumptions on the spacetime symmetry. At order zero in derivatives, the current has to be constant, $J^\mu_\alpha=e^\mu_\alpha$. In this case, eqs.~\eqref{currentconservation} and~\eqref{currenttransfo} impose the same condition on the coupling, namely $e^\mu_\alpha f^\alpha_{\beta\gamma}=0$. Integrating the current according to eq.~\eqref{CSmaster} then recovers the CS term shown in eq.~\eqref{LOlagrangian}, while the special case of rotational invariance is displayed as $\La^{(1)}_\text{CS}$ in eq.~\eqref{CSterms}. At order one, there is no covariant current since the simplest building block we have, $G^\alpha_{\mu\nu}$, is already of order two. This explains in very elementary terms why the part of the effective Lagrangian with two derivatives is strictly gauge-invariant.

At order two, the current has to be proportional to the field-strength tensor,
\begin{equation}
J^\mu_\alpha=c^{\mu\nu\lambda}_{\alpha\beta}G^\beta_{\nu\lambda}.
\end{equation}
As explained above, the covariance of the current requires that the coupling $c^{\mu\nu\lambda}_{\alpha\beta}$ is an invariant tensor of $H$. Also, it can be without loss of generality assumed antisymmetric in the indices $\nu,\lambda$. The conservation condition~\eqref{currentconservation} on the other hand takes the form
\begin{equation}
0=c^{\mu\nu\lambda}_{\alpha\beta}D_\mu G^\beta_{\nu\lambda}=2c^{\mu\nu\lambda}_{\alpha\beta}\de_\mu\de_\nu B^\beta_\lambda+\dotsb,
\end{equation}
where the ellipsis denotes terms with fewer than two derivatives acting on $B^\alpha_\mu$. Since all components of $B^\alpha_\mu$ with different $\alpha$ and $\mu$ are in principle independent functions of spacetime, this implies that $c^{\mu\nu\lambda}_{\alpha\beta}$ has to be antisymmetric in $\mu,\nu$. Given the assumed antisymmetry in $\nu,\lambda$, it must thus be fully antisymmetric in all three indices $\mu,\nu,\lambda$. This is a necessary as well as sufficient condition for current conservation, since the whole $c^{\mu\nu\lambda}_{\alpha\beta}D_\mu G^\beta_{\nu\lambda}$ then vanishes by means of the Bianchi identity. Integration using eq.~\eqref{CSmaster} now finally gives
\begin{equation}
\La_\text{CS}^{(3)}=c^{\lambda\mu\nu}_{\alpha\beta}B^\alpha_\lambda(\de_\mu B^\beta_\nu+\tfrac13f^\beta_{\gamma\delta}B^\gamma_\mu B^\delta_\nu).
\end{equation}
The coupling $c^{\lambda\mu\nu}_{\alpha\beta}$ is fully antisymmetric in $\lambda,\mu,\nu$ and is an invariant tensor of the unbroken subgroup $H$. Also, it can without loss of generality be considered symmetric in $\alpha,\beta$, since swapping these two indices changes the Lagrangian at most by a surface term.

In one spatial dimension, a rank-three fully antisymmetric tensor does not exist, while in two dimensions, $c^{\lambda\mu\nu}_{\alpha\beta}$ has to be proportional to $\epsilon^{\lambda\mu\nu}$. Finally, in three dimensions, it can be equivalently written in terms of the dual vector, $c^{\lambda\mu\nu}_{\alpha\beta}=c_{\kappa,\alpha\beta}\epsilon^{\kappa\lambda\mu\nu}$. In principle, the internal group structure of the vector $c_{\kappa,\alpha\beta}$ can be chosen independently for each component $\kappa$. Under the assumption of rotational invariance, only the $\kappa=0$ component can be nonzero, while in higher dimensions no rank-three fully antisymmetric and rotationally invariant tensor exists. This reproduces the most general rotationally invariant CS Lagrangian of order three in derivatives, given in eq.~\eqref{CSterms}. Note that our result is implicit in older works dealing with the general problem of classification of effective actions without the assumption of gauge invariance~\cite{DHoker:1994ti,DHoker:1995it,deAzcarraga:1997gn}. It is therefore worth emphasizing that we have obtained it using solely elementary field theory. In the next subsection, we discuss its implications in more detail.

So far, we have found nontrivial CS terms at orders one and three in derivatives. It can be shown that no such terms appear at order four, the highest order of concern in this paper. Although a proof of this statement is elementary, it is rather lengthy and the details are therefore deferred to appendix~\ref{app:CSfour}. Before concluding, it is, however, instructive to inspect the variation of the CS Lagrangians under the gauge transformation~\eqref{Bphitransfo}. For the order-one term, this is nearly trivial,
\begin{equation}
\delta\La^{(1)}_\text{CS}=e^\mu_\alpha(\de_\mu\epsilon^\alpha+f^\alpha_{\beta\gamma}B^\beta_\mu\epsilon^\gamma)=\de_\mu(e^\mu_\alpha\epsilon^\alpha),
\end{equation}
where the second term drops out as a consequence of the $H$-invariance of $e^\mu_\alpha$. Trivial as it seems, it is good to realize that the surface term actually depends nontrivially on the NG fields $\pi^a$. Indeed, as is clear from eq.~\eqref{defk}, the parameter of the compensating transformation $\h\in H$ by which the auxiliary field $B^\alpha_\mu$ shifts equals $\epsilon^\alpha=\epsilon^ik^\alpha_i(\pi)$, where $\epsilon^i$ is the parameter of the original gauge transformation, $\g\in G$. Finally, we just add that the order-three CS term varies by
\begin{equation}
\delta\La^{(3)}_\text{CS}=\de_\lambda(c^{\lambda\mu\nu}_{\alpha\beta}\epsilon^\alpha\de_\mu B^\beta_\nu).
\end{equation}
The proof of this statement follows upon a brief manipulation using the invariance condition on $c^{\lambda\mu\nu}_{\alpha\beta}$ and the Jacobi identity for the structure constants.


\subsection{Physical content of the Chern-Simons terms}
\label{subsec:CSphysical}

The CS terms are singled out by our construction, based on the auxiliary field variables $\phi^a_\mu$ and $B^\alpha_\mu$. A natural question then arises, whether this division is purely technical or whether the CS terms differ from the invariant part of the Lagrangian $\La_\text{inv}$ in actual physical consequences. One aspect of the CS terms certainly is special: as a consequence of the global topology of the coset space $G/H$, the couplings $e_\alpha$ and $c_{\alpha\beta}$ are, as a rule, quantized. This follows from a rather deep analogy of our CS terms with the so-called Wess-Zumino term in $\chi$PT~\cite{Witten:1983tw,Altland:book}, discussed in detail in the companion paper~\cite{CSpaper}. The topological nature of the order-one term $\La^{(1)}_\text{CS}$, for instance, manifests itself in the Berry phase that the ground state of the system acquires when adiabatically dragged through $G/H$ by a weak external field~\cite{Watanabe:2014fva,Aitchison:1986qn}.

Since our concern in this paper is the construction of EFTs for NG bosons, we now concentrate on the perturbative interactions induced by the CS terms. The case of $\La^{(1)}_\text{CS}$ was already discussed in detail in section~\ref{subsec:LOphysical}. We saw in eq.~\eqref{LOlagrangianexpansion} that this CS term affects the spectrum of NG bosons by canonically conjugating some of the fields $\pi^a$, provided two conditions are satisfied: $e_\alpha\neq0$ for some $\alpha$ and the existence of generators $T_a,T_b$ such that $f^\alpha_{ab}\neq0$. The first condition amounts to the presence of unbroken charge density in the ground state, and can only be fulfilled when $T_\alpha$ (or more precisely $T^\alpha$, see appendix~\ref{app:tensor} for the definition of the notation) generates a $\gr{U(1)}$ factor of $H$. The second condition guarantees that $T_\alpha$ does not commute with the whole group $G$, in which case the charge density would completely decouple from the dynamics of the NG bosons~\cite{Leutwyler:1993gf}.

Let us reformulate the latter condition in a more formal fashion which will prove useful below when discussing $\La_\text{CS}^{(3)}$. \emph{Assume} that there is a set of couplings $E^\mu_i$ such that $E^\mu_if^i_{jk}=0$ and $E^\mu_\alpha=e^\mu_\alpha$. This means that $\La^{(1)}_\text{CS}$, expressed in terms of $B^\alpha_\mu$ as in eq.~\eqref{CSterms}, can be embedded into a CS term for the entire field $\tilde A_\mu=\gtr{U^{-1}}A_\mu=\phi_\mu+B_\mu$, $\tilde\La_\text{CS}^{(1)}=E^\mu_i\tilde A^i_\mu$. Since $\tilde A^i_\mu$ differs from the original gauge field $A^i_\mu$ just by a gauge transformation, the action stays unchanged if we replace $\tilde A^i_\mu$ by $A^i_\mu$.\footnote{Strictly speaking, this is only true for topologically trivial NG field configurations. In general, the action may shift by a topological $\theta$-term, which nevertheless does not qualitatively modify our argument~\cite{CSpaper}.} On the other hand, we obviously have
\begin{equation}
\tilde\La_\text{CS}^{(1)}=E^\mu_\alpha B^\alpha_\mu+E^\mu_a\phi^a_\mu=\La_\text{CS}^{(1)}+E^\mu_a\phi^a_\mu.
\end{equation}
Our assumption on $E^\mu_i$ implies as a special case that $E^\mu_af^a_{\alpha b}=0$. Hence $E^\mu_a$ is an invariant tensor of $H$ and the term $E^\mu_a\phi^a_\mu$ can be absorbed into the redefinition of $\La_\text{inv}$. We have therefore established that provided the couplings $E^\mu_i$ exist, $\La_\text{CS}^{(1)}$ is actually independent of the NG fields, being gauge-equivalent to $E^\mu_iA^i_\mu$ plus a term that belongs to $\La_\text{inv}$. In rotationally invariant systems, $E^\mu_i=E_i\delta^{\mu0}$, and the couplings $E_i$ with the desired properties exist when the Lie algebra of $G$ possesses a $\gr{U(1)}$ generator $E^iT_i$ which, when projected to the Lie subalgebra of $H$, reduces to $e^\alpha T_\alpha$.

The same steps can be followed in case of $\La_\text{CS}^{(3)}$. We first \emph{assume} that there is a set of couplings $C^{\lambda\mu\nu}_{ij}$ such that $C^{\lambda\mu\nu}_{\alpha\beta}=c^{\lambda\mu\nu}_{\alpha\beta}$ and $C^{\lambda\mu\nu}_{\ell j}f^\ell_{ik}+C^{\lambda\mu\nu}_{i\ell}f^\ell_{jk}=0$. We use this to promote $\La_\text{CS}^{(3)}$ to a CS term containing the full gauge field $\tilde A^i_\mu$, $\tilde\La_\text{CS}^{(3)}=C^{\lambda\mu\nu}_{ij}\tilde A^i_\lambda(\de_\mu\tilde A^j_\nu+\tfrac13f^j_{k\ell}\tilde A^k_\mu\tilde A^\ell_\nu)$. By an explicit manipulation, we can then show that $\La_\text{CS}^{(3)}$ is gauge-equivalent to
\begin{equation}
C^{\lambda\mu\nu}_{ij}A^i_\lambda\Bigl(\de_\mu A^j_\nu+\frac13f^j_{k\ell}A^k_\mu A^\ell_\nu\Bigr)-C^{\lambda\mu\nu}_{a\alpha}\phi^a_\lambda G^\alpha_{\mu\nu}-C^{\lambda\mu\nu}_{ab}\phi^a_\lambda D_\mu\phi^b_\nu-\frac13C^{\lambda\mu\nu}_{ai}f^i_{bc}\phi^a_\lambda\phi^b_\mu\phi^c_\nu.
\label{longidentity}
\end{equation}
In other words, $\La_\text{CS}^{(3)}$ is independent of the NG fields up to terms that can be absorbed into a redefinition of the couplings in $\La_\text{inv}$.

Under what conditions does the extension $C_{ij}$ exist? (For the sake of simplicity, we drop the Lorentz indices here since they are not essential for the discussion.) The coupling $c_{\alpha\beta}$ defines an $H$-invariant symmetric bilinear form on the Lie algebra of $H$, and mathematically speaking we are therefore investigating the existence of its extension to a $G$-invariant symmetric bilinear form on $G$. According to appendix~\ref{app:tensor}, $c^\alpha_{\phantom\alpha\beta}$ commutes with all generators of $H$ in the adjoint representation. Provided that $H$ is simple, its adjoint representation is irreducible, and by Schur's lemma $c^\alpha_{\phantom\alpha\beta}$ has to be proportional to $\delta^\alpha_\beta$. In this case, it can be naturally extended to $C^i_{\phantom ij}\propto\delta^i_j$. When $H$ is not simple, $c^\alpha_{\phantom\alpha\beta}$ is allowed by Schur's lemma to contain several blocks, each proportional to unity, but with different eigenvalues. In case two such blocks lie in the same invariant subspace of $G$ under the adjoint action, there is obviously no $G$-invariant extension of $c^\alpha_{\phantom\alpha\beta}$.

We conclude that the necessary condition for $\La_\text{CS}^{(3)}$ to trigger interactions among NG bosons is that $H$ is not simple. As a concrete example, consider the symmetry-breaking pattern $\gr{SU(2)\times SU(2)\to U(1)\times U(1)}$. The adjoint action of $G$ splits the generators into two invariant spaces, one for each of the $\gr{SU(2)}$ factors.  Consequently, the most general $G$-invariant bilinear form has to be proportional to unity on either of them. Denoting the indices of the two $\gr{SU(2)}$s as $\underline{i},\underline{j},\dotsc$ and $\overline{i},\overline{j},\dotsc$, this means that
\begin{equation}
C^{\underline i}_{\phantom i\underline j}\propto\delta^i_j,\qquad C^{\overline i}_{\phantom i\overline j}\propto\delta^i_j,\qquad C^{\underline{i}}_{\phantom i\overline{j}}=C^{\overline{i}}_{\phantom i\underline{j}}=0.
\end{equation}
However, the unbroken subgroup $H=\gr{U(1)\times U(1)}$, assumed without loss of generality to correspond to $T_{\underline{3}}$ and $T_{\overline{3}}$, is Abelian so that $c_{\alpha\beta}$ can be chosen completely arbitrarily. Once we choose nonzero $c_{\underline{3}\overline{3}}=c_{\overline{3}\underline{3}}$, an extension of $c_{\alpha\beta}$ to a $G$-invariant coupling obviously does not exist. We can go even further and say that the parts of $\La_\text{CS}^{(3)}$ corresponding to $c_{\underline{3}\underline{3}}$ and $c_{\overline{3}\overline{3}}$ can be absorbed into the redefinition of $\La_\text{inv}$ via eq.~\eqref{longidentity}, while that containing $c_{\underline{3}\overline{3}}=c_{\overline{3}\underline{3}}$ will give nontrivial interactions among the NG bosons. What kind of interactions does it then represent? Note that since $G$ is given by a direct product of two subgroups, the fields $\phi^a_\mu$, $B^\alpha_\mu$ split into completely separated parts carrying the NG and gauge fields of the respective $\gr{SU(2)}$ factors. A glance at eqs.~\eqref{LOoperators} and~\eqref{listofoperators} reveals that up to order three in derivatives, there is no term in the invariant Lagrangian the would mix the fields from the two subgroups. Therefore, $\La_\text{CS}^{(3)}$ in this case provides the \emph{leading} interaction among all the NG bosons of the theory.

To see the above more explicitly, let us for simplicity discard the external gauge fields $A^i_\mu$. Using eq.~\eqref{auxfields} together with the expansions~\eqref{expparam}, it is easy to obtain the leading term in the power expansion of $\La_\text{CS}^{(3)}$,
\begin{equation}
\La_\text{CS}^{(3)}=\tfrac14c^{\lambda\mu\nu}_{\alpha\beta}f^\alpha_{ab}f^\beta_{cd}\pi^a\de_\lambda\pi^b\de_\mu\pi^c\de_\nu\pi^d+\dotsb.
\label{WZW}
\end{equation}
This interaction term is fully antisymmetric in the indices $a,b,c,d$, and thus requires that the dimension of $G/H$ is at least four. Our $\gr{[SU(2)\times SU(2)]/[U(1)\times U(1)]}$ example therefore provides a minimal model in which $\La_\text{CS}^{(3)}$ is nontrivial, another simple example being for instance the symmetry-breaking pattern $\gr{SU(3)\to SU(2)\times U(1)}$. Finally, note that the interaction~\eqref{WZW} has a very similar structure to the Wess--Zumino term in $\chi$PT~\cite{Scherer:2002tk}. The important difference between the two however is that our interaction arises from a strictly \emph{gauge-invariant} action.


\subsection{Explicit symmetry breaking}
\label{subsec:explicitSB}

So far, we have assumed that the symmetry of the physical system under the group $G$ is exact. However, examples of exact \emph{global} symmetries are rare, and as a rule correspond to Abelian groups. In realistic situations, non-Abelian global symmetries are broken explicitly, albeit weakly. A prototypical example is the chiral symmetry of QCD, which is explicitly broken by the nonzero masses of the quarks. It is therefore mandatory to understand how to incorporate the effects of such explicit symmetry breaking in the EFT.

Following the treatment of ref.~\cite{Leutwyler:1993iq}, we assume that in the microscopic theory, the $G$-invariance is broken by a term in the Lagrangian, $m_\sigma\mathcal O^\sigma$, containing a set of operators that transform in some (nontrivial) linear representation of $G$: $\mathcal O^\sigma\xrightarrow{\g}\mathcal{O}'^\sigma=D(\g)^\sigma_{\phantom{\sigma}\rho}\mathcal{O}^\rho$. The invariance under $G$ can be formally restored by assigning the parameters $m_\sigma$ a contragredient transformation rule, $m'_\sigma=D(\g^{-1})^\rho_{\phantom{\rho}\sigma}m_\rho$. In the EFT, the parameters $m_\sigma$ can be treated as a background field with the above transformation rule. The generating functional $\Gamma\{A,m\}$, defined in analogy with eq.~\eqref{genfunc}, is invariant under a simultaneous gauge transformation of the fields $A^i_\mu$ and $m_\sigma$. The invariance theorem~\ref{thm:invariancetheorem} can be seen to still hold in this case so that the effective action $S_\text{eff}\{\pi,A,m\}$ is gauge-invariant under a simultaneous transformation of all its arguments.

Upon the gauge transformation that eliminates the NG fields from the EFT, we find, analogously to eq.~\eqref{gaugeinvartrick},
\begin{equation}
S_\text{eff}\{\pi,A,m\}=S_\text{eff}\{0,\gtr{U(\pi)^{-1}}A,\Xi(\pi)\},
\end{equation}
where $\Xi_\sigma(\pi)=D(U(\pi))^\rho_{\phantom\rho\sigma}m_\rho$. The dependence of the action on the fields $A^i_\mu$ and $m_\sigma$ at $\pi=0$ is constrained by the invariance under the unbroken subgroup $H$. Vice versa, repeating the steps in eq.~\eqref{proofgaugetrick}, it is easy to see that every $H$-invariant functional $F\{A,m\}$ gives rise to a $G$-invariant action by means of $S_\text{eff}\{\pi,A,m\}=F\{\gtr{U(\pi)^{-1}}A,\Xi(\pi)\}$. The most general effective Lagrangian respecting all the symmetries is therefore obtained using three basic building blocks: the already familiar fields $\phi^a_\mu$ and $B^\alpha_\mu$, and $\Xi_\sigma$. By an extension of the proof of theorem~\ref{thm:Laginv}, one can likewise show that the full dependence of the Lagrangian on the field $\Xi_\sigma$ resides in its invariant part, $\La_\text{inv}[\phi,B,\Xi]$.

The above argument allows us to determine the dependence of the effective Lagrangian on the explicit-symmetry-breaking parameters using the same strategy as before, namely by listing all covariant operators up to the desired order in the derivative expansion and contracting their indices with $H$-invariant tensor couplings. The precise way that $\Xi_\sigma$, and hence $m_\sigma$, appears in the Lagrangian depends on how we \emph{define} its power counting. Since we expect $m_\sigma$ to give rise to a quadratic mass term for the NG modes, it is natural to count it as second-order in the derivative expansion. Up to order four in derivatives, we then have the following operators containing at least one factor of $\Xi_\sigma$:
\begin{equation}
\begin{split}
\text{order 2: }&\Xi_\sigma.\\
\text{order 3: }&\Xi_\sigma\phi^a_\mu,\xcancel{D_\mu\Xi_\sigma}.\\
\text{order 4: }&\Xi_\rho\Xi_\sigma,\Xi_\sigma\phi^a_\mu\phi^b_\nu,\Xi_\sigma D_\mu\phi^a_\nu,\xcancel{D_\nu\Xi_\sigma\phi^a_\mu},\xcancel{D_\mu D_\nu\Xi_\sigma},\Xi_\sigma G^\alpha_{\mu\nu}.
\end{split}
\end{equation}
We again crossed out the operators that are either total derivatives or can be expressed in terms of the others by partial integration. Working out all the possible contractions of the spacetime indices in these operators, allowed by the spacetime symmetry, reproduces the result advertised in section~\ref{subsubsec:Lsb}.


\section{Conclusions}
\label{sec:conclusions}

In this paper, we have worked out a systematic framework for the construction of effective actions for the NG bosons of a spontaneously broken internal symmetry in arbitrary quantum many-body systems. Building on the pioneering work of ref.~\cite{Leutwyler:1993iq}, we have provided explicit expressions for the most general effective Lagrangian up to order four in the gradient expansion. Although we have assumed rotational invariance, the generalization of the results to other spacetime symmetries is in principle fully straightforward. To conclude the paper, we would like to make a number of comments on our results.

First, we worked out the constraints on the effective Lagrangian based solely on the continuous symmetry. Real physical systems often possess additional, discrete symmetries such as parity, time reversal or charge conjugation. These may dramatically reduce the number of independent parameters in the Lagrangian, as we saw in the examples discussed in section~\ref{sec:examples}.

Second, from the outset we made the assumption that NG bosons are the only low-energy degrees of freedom. This is often not the case, a generic example being the gapless fermionic excitations in presence of a Fermi sea. In principle, adding such non-NG degrees of freedom is straightforward using the techniques developed in refs.~\cite{Coleman:1969sm,Callan:1969sn}.

Last, for the sake of simplicity, we assumed that the continuous symmetry that defines the EFT can be gauged, that is, there are no anomalies. Taking the anomalies into account is nontrivial, but in principle feasible~\cite{Leutwyler:1993iq}. One has to construct a contribution to the effective action that reproduces the anomaly in the Ward-Takahashi identities for the Green's functions of the microscopic theory. Once this is found, the remaining part of the effective action can be obtained using the methods presented in this paper.

Of course, constructing the effective action is just the initial step in a more long-term program. To get a full-fledged EFT framework, our results need to be augmented with tools for performing loop computations in the many-body system in question. These have already been developed for numerous concrete examples, and will be invaluable for the applications of the formalism presented here. To work out such explicit cases will constitute the main direction of our future efforts.


\acknowledgments

We gratefully appreciate discussions and correspondence with Paulo Bedaque, Heinrich Leutwyler, Sergej Moroz, Mikko Sainio, Haruki Watanabe, and Uwe-Jens Wiese. The work of T.B.~and A.V.~was supported by the Academy of Finland, grant No.~273545, and the work of T.B.~in addition by the Austrian Science Fund (FWF), grant No.~M 1603-N27. C.P.H.~was supported through the project ``Redes Tematicas de Colaboracion Academica 2013, UCOL-CA-56''. We furthermore acknowledge the hospitality of IFT UAM-CSIC, afforded through the Centro de Excelencia Severo Ochoa Program under grant SEV-2012-0249, which was instrumental for the establishment of our collaboration.


\appendix


\section{Invariant tensor couplings}
\label{app:tensor}

The effective couplings that appear in the effective Lagrangian are invariant tensors of the unbroken subgroup $H$. They carry indices of the representations $R_\phi$ and $R_B$ of $H$ in which the fields $\phi^a_\mu$ and $B^\alpha_\mu$ transform (the latter being the adjoint representation). Mathematically, finding all independent parameters contained in the effective Lagrangian therefore corresponds to finding all singlets of $H$ in tensor products of the appropriate number of $R_\phi$s and $R_B$s, corresponding to the operator that one deals with. Obviously, the actual number of independent parameters strongly depends on the size and structure of the unbroken subgroup $H$. (In the unfortunate case that the continuous symmetry is fully broken, the couplings can take completely arbitrary values.) While the general solution to this problem is probably available in the mathematical literature, for practical purposes it is most convenient to find it case by case using tensor methods~\cite{Cvitanovic}.

Here we discuss the consequences of the invariance constraint~\eqref{invariancecondition} in the simplest cases that occur repeatedly throughout our analysis. Let us first introduce the necessary notation. Choose some faithful representation for the generators $T_i$ and define a bilinear form on the Lie algebra of $G$ by $\Delta_{ij}=\tr(T_iT_j)$.\footnote{Our effective Lagrangian of course depends solely on the structure constants of $G$ and hence is independent of such a choice of representation.} It is assumed to be nondegenerate, but not necessarily proportional to $\delta_{ij}$. The latter can always be ensured for compact semisimple Lie algebras by a suitable choice of basis.  The matrix $\Delta_{ij}$ and its inverse $\Delta^{ij}$ can be used to lower and raise indices of covariantly transforming objects. 

The matrix $\Delta_{ij}$ represents a symmetric rank-two tensor and is obviously invariant under the adjoint action of $G$, which implies a condition of the same type as~\eqref{invariancecondition},
\begin{equation}
\Delta_{\ell j}f^\ell_{ki}+\Delta_{i\ell}f^\ell_{kj}=0.
\label{Deltainv}
\end{equation}
This can also be proven directly from the cyclicity of the trace, $\tr([T_i,T_k]T_j)=\tr(T_i[T_k,T_j])$. It follows immediately that the combination $\Delta_{i\ell}f^\ell_{jk}$ is fully antisymmetric in the indices $i,j,k$. By restoring the proper ordering of indices in the structure constants, $f^i_{jk}=f^i_{\phantom ijk}$, and setting $f_{ijk}=\Delta_{i\ell}f^\ell_{\phantom\ell jk}$, this antisymmetry can be interpreted as a generalization of the usual property of the structure constants in a basis-independent way. Likewise, it is easy to show that the tensor $f^k_{\phantom kij}\Delta^{j\ell}=f^{k\phantom i\ell}_{\phantom ki}$ is antisymmetric in the indices $k,\ell$.

In the following, we will in addition assume that $\Delta_{ij}$ is block-diagonal with respect to the broken and unbroken indices, that is, $\Delta_{a\alpha}=\Delta_{\alpha a}=0$. This is reasonable: the broken generators are defined to be ``orthogonal'' to the unbroken ones. More precisely, a specific choice of indices in eq.~\eqref{Deltainv} gives (we use the fact that the algebra of unbroken generators always closes so that $f^a_{\alpha\beta}=0$)
\begin{equation}
0=\Delta_{\gamma a}f^\gamma_{\alpha\beta}=\Delta_{ia}f^i_{\alpha\beta}=-\Delta_{\beta i}f^i_{\alpha a}=-\Delta_{\beta\gamma}f^\gamma_{\alpha a}.
\end{equation}
Since $\Delta_{ij}$, and thus $\Delta_{\alpha\beta}$, is by assumption nondegenerate, we can divide by it and thereby obtain $f^\gamma_{\alpha a}=0$. This is nothing but our assumption that the broken generators furnish a representation, $R_\phi$, of the unbroken subgroup $H$. A practical consequence of the assumed block-diagonal structure of $\Delta_{ij}$ is that (un)broken indices remain (un)broken after raising or lowering so that, for instance, $u_\alpha v^\alpha=u^\alpha v_\alpha$. This would not necessarily hold otherwise.


\subsection{Couplings of the type $c_\alpha$ and $c_a$}
\label{subapp:ca}

The invariance conditions here read $c_\alpha f^\alpha_{\beta\gamma}=0$ and $c_af^a_{\alpha b}=0$, respectively. A simple manipulation using the above-defined symmetric form $\Delta_{ij}$ leads to
\begin{equation}
0=c_\alpha f^\alpha_{\beta\gamma}\Delta^{\gamma\delta}T_\delta=-c_\alpha f^\delta_{\beta\gamma}\Delta^{\gamma\alpha}T_\delta=\imag c^\gamma[T_\beta,T_\gamma]=\imag[T_\beta,c^\gamma T_\gamma].
\end{equation}
(In the first step, we used the antisymmetry of $f^\alpha_{\beta\gamma}\Delta^{\gamma\delta}$ in the upper two indices.) We conclude that the $H$-invariance of $c_\alpha$ is equivalent to the statement that the matrix $c^\alpha T_\alpha$ commutes with all unbroken generators. Likewise, $c_a$ is an invariant tensor of $H$ if and only if $c^aT_a$ commutes with all unbroken generators. Thus, the possible existence of a rank-one invariant tensor of $H$ is determined by group theory: the space of (un)broken generators must contain a singlet of $H$. In particular, the couplings $e_i$ in $\La^{(0,1)}_\text{eff}$ correspond to the vacuum expectation values of the generators $T_i$. The group-theoretic condition on $e_i$ translates into the (obvious) statement that only generators commuting with the unbroken subgroup $H$ can have a nonzero vacuum expectation value. 


\subsection{Couplings of the type $c_{\alpha\beta}$ and $c_{ab}$}
\label{subapp:cab}

The invariant tensor $c_{\alpha\beta}$ satisfies a condition analogous to eq.~\eqref{Deltainv}. Raising the first index of $c_{\alpha\beta}$ with $\Delta^{\alpha\beta}$, it can be rewritten as
\begin{equation}
c^\alpha_{\phantom\alpha\gamma}f^\gamma_{\delta\beta}-f^\alpha_{\delta\gamma}c^\gamma_{\phantom\gamma\beta}=0.
\end{equation}
In other words, the matrix $c^\alpha_{\phantom\alpha\beta}$ commutes with all generators of $H$ in the adjoint representation. This determines $c^\alpha_{\phantom\alpha\beta}$ completely up to a few unknown parameters. When $H$ is simple, its adjoint representation is irreducible and by Schur's lemma, $c^\alpha_{\phantom\alpha\beta}$ must equal $\delta^\alpha_\beta$ up to an overall factor. A general compact group $H$ is given by a product $H_1\times\dotsb\times H_p\times\gr{U(1)}^q$ (possibly multiplied by another, discrete group), where all $H_i$ are simple. Then, $c^\alpha_{\phantom\alpha\beta}$ is determined by a single constant on every simple factor $H_i$, while it can take arbitrary values on the Abelian part $\gr{U(1)}^q$.

A completely analogous statement holds for $c_{ab}$, except that now $c^a_{\phantom ab}$ commutes with all generators of $H$ in the representation $R_\phi$. The allowed values of  $c^a_{\phantom ab}$ are determined by the decomposition of $R_\phi$ into irreducible components. By Schur's lemma, $c^a_{\phantom ab}$ has to be proportional to $\delta^a_b$ on every irreducible representation which appears only once in the decomposition of $R_\phi$. When $R_\phi$ itself is irreducible, $c_{ab}$ must be proportional to $\Delta_{ab}$ and so is necessarily symmetric. This in particular means that the antisymmetric two-derivative term $\epsilon^{rs}\phi^a_r\phi^b_s$ in the leading-order Lagrangian~\eqref{LOlagrangian}, allowed in two-spatial dimensions by rotational invariance, is forbidden by the internal symmetry.


\section{The $c_7$ operator in chiral perturbation theory}
\label{app:c7}

Here we provide the missing details behind the construction of the effective Lagrangian for $\chi$PT, worked out in section~\ref{subsec:pions}. Our aim is to analyze the consequences of the operator $c_7$ in eq.~\eqref{order4naive}. Using eq.~\eqref{DSigma} for $D_\mu\Phi_\nu$ and the identity
\begin{equation}
(D_\mu D_\nu\CPT)\CPT^{-1}+\CPT(D_\mu D_\nu\CPT^{-1})=-(D_\mu\CPT D_\nu\CPT^{-1}+D_\nu\CPT D_\mu\CPT^{-1}),
\label{auxCPT}
\end{equation}
which follows by differentiating the relation $\CPT\CPT^{-1}=\openone$ twice, we immediately get
\begin{equation}
\begin{split}
\tr(D_\mu\Phi_\nu D^\mu\Phi^\nu)={}&-\frac18\tr\bigl[D_\mu\CPT D^\mu\CPT^{-1}D_\nu\CPT D^\nu\CPT^{-1}+D_\mu\CPT D_\nu\CPT^{-1}D^\mu\CPT D^\nu\CPT^{-1}\\
&-2(D_\mu D_\nu\CPT)(D^\mu D^\nu\CPT^{-1})\bigr].
\end{split}
\end{equation}
Consequently, modulo terms that can be absorbed into a redefinition of the couplings $c_1$ and $c_2$, the $c_7$ operator is proportional to $\tr[(D_\mu D_\nu\CPT)(D^\mu D^\nu\CPT^{-1})]$. We therefore focus on this combination of the fields.

Let us now make a step aside and rewrite the equation of motion $D_\mu\Phi^\mu=0$, eq.~\eqref{EOMLO}, in terms of $\CPT$. Using eq.~\eqref{DSigma}, it takes the form
\begin{equation}
D_\mu D^\mu\CPT=\CPT(D_\mu D^\mu\CPT^{-1})\CPT.
\end{equation}
Substituting for $D_\mu D^\mu\CPT^{-1}$ from eq.~\eqref{auxCPT}, this becomes $D_\mu D^\mu\CPT=(D_\mu\CPT)\CPT^{-1}(D^\mu\CPT)$, and equivalently $D_\mu D^\mu\CPT^{-1}=(D_\mu\CPT^{-1})\CPT(D^\mu\CPT^{-1})$. As an immediate consequence, we get that $\tr[(D_\mu D^\mu\CPT)(D_\nu D^\nu\CPT^{-1})]=\tr(D_\mu\CPT D^\mu\CPT^{-1}D_\nu\CPT D^\nu\CPT^{-1})$ for fields satisfying the equation of motion. This is used in the next step, where we rewrite a similar operator using integration by parts (the resulting equivalence up to a surface term is indicated by the symbol $\sim$),
\begin{align}
\notag
\tr[(D_\mu D_\nu\CPT)(D^\nu D^\mu\CPT^{-1})]\sim{}&-\tr[(D_\nu D_\mu D^\nu\CPT)(D^\mu\CPT^{-1})]\\
\notag
\sim{}&-\tr\bigl\{([D_\nu,D_\mu]D^\nu\CPT)(D^\mu\CPT^{-1})-(D_\nu D^\nu\CPT)(D_\mu D^\mu\CPT^{-1})\bigr\}\\
\label{auxCPT2}
={}&\imag\tr(F^L_{\mu\nu}D^\mu\CPT D^\nu\CPT^{-1}+F^R_{\mu\nu}D^\mu\CPT^{-1}D^\nu\CPT)\\
\notag
&+\tr[(D_\mu D^\mu\CPT)(D_\nu D^\nu\CPT^{-1})].
\end{align}
We also used that a commutator of two covariant derivatives gives the field-strength tensor,
\begin{equation}
[D_\mu,D_\nu]\CPT=-\imag F^L_{\mu\nu}\CPT+\imag\CPT F^R_{\mu\nu}.
\label{auxCPT3}
\end{equation}
Since the first term on the right-hand side of eq.~\eqref{auxCPT2} arises already from the $c_9$ and $c_{11}$ operators, see eq.~\eqref{c9c11}, while the second term from $c_1$, as argued above, we conclude that upon using the equation of motion, a term of the type $\tr[(D_\mu D_\nu\CPT)(D^\nu D^\mu\CPT^{-1})]$ can be completely dropped from the Lagrangian.

This seemingly irrelevant observation allows us to antisymmetrize the Lorentz indices in $\tr[(D_\mu D_\nu\CPT)(D^\mu D^\nu\CPT^{-1})]$, which is what we are eventually after. We thus obtain
\begin{equation}
\tr\bigl\{([D_\mu,D_\nu]\CPT)([D^\mu,D^\nu]\CPT^{-1})\bigr\}=\tr(F^L_{\mu\nu}F^{L\mu\nu}+F^R_{\mu\nu}F^{R\mu\nu}-2F^L_{\mu\nu}\CPT F^{R\mu\nu}\CPT^{-1}).
\end{equation}
This ultimately confirms that, up to terms that vanish upon using the equation of motion and terms that can be absorbed into a redefinition of the other couplings present in the Lagrangian~\eqref{order4naive}, the $c_7$ operator reduces to the first line of eq.~\eqref{c9c11} with a flipped sign in front of $2F^L_{\mu\nu}\CPT F^{R\mu\nu}\CPT^{-1}$. Together, these operators therefore give rise to the independent couplings $\tilde  c_5$ and $\tilde c_6$ in our Lagrangian~\eqref{order4final}.


\section{Covariance of the scalar and vector currents}
\label{app:covariance}

In this appendix, we discuss the transformation properties of the scalar and vector currents, defined by a functional derivative of the action with respect to the fields $\phi^a_\mu$ and $B^\alpha_\mu$,
\begin{equation}
\Sigma^\mu_a[\phi,B]=\frac{\delta S\{\phi,B\}}{\delta\phi^a_\mu},\qquad J^\mu_\alpha[\phi,B]=\frac{\delta S\{\phi,B\}}{\delta B^\alpha_\mu}.
\end{equation}
Consider an arbitrary infinitesimal shift of the fields, $\phi^a_\mu\to\phi^a_\mu+u^a_\mu$ and $B^\alpha_\mu\to B^\alpha_\mu+v^\alpha_\mu$, and the induced shift of the action,
\begin{equation}
S\{\phi+u,B+v\}-S\{\phi,B\}=\int\dd x\,(u^a_\mu\Sigma^\mu_a[\phi,B]+v^\alpha_\mu J^\mu_\alpha[\phi,B]).
\label{auxC1}
\end{equation}
Next apply the gauge transformation~\eqref{Bphitransfo} to this equation. This leads to
\begin{equation}
\begin{split}
\int\dd x\,(u^a_\mu\Sigma^\mu_a[\phi',B']+v^\alpha_\mu J^\mu_\alpha[\phi',B'])&=S\{\phi'+u,B'+v\}-S\{\phi',B'\}\\
&=S\{\phi'+u,B'+v\}-S\{\phi,B\}.
\end{split}
\label{auxC2}
\end{equation}
Taking the difference of eqs.~\eqref{auxC2} and~\eqref{auxC1} allows us to determine the transformation of the currents, induced by the gauge transformation of the fields~\eqref{Bphitransfo},
\begin{equation}
\begin{split}
\int\dd x\,&(u^a_\mu\delta\Sigma^\mu_a[\phi,B]+v^\alpha_\mu\delta J^\mu_\alpha[\phi,B])=S\{\phi'+u,B'+v\}-S\{\phi+u,B+v\}\\
&=\int\dd x\,\bigl\{f^a_{b\alpha}\phi^b_\mu\epsilon^\alpha\Sigma^\mu_a[\phi+u,B+v]+(f^\alpha_{\beta\gamma}B^\beta_\mu\epsilon^\gamma+\de_\mu\epsilon^\alpha)J^\mu_\alpha[\phi+u,B+v]\bigr\}\\
&\approx-\int\dd x\,\bigl(f^b_{a\alpha}u^a_\mu\epsilon^\alpha\Sigma^\mu_b[\phi,B]+f^\gamma_{\alpha\beta}v^\alpha_\mu\epsilon^\beta J^\mu_\gamma[\phi,B]\bigr).
\end{split}
\end{equation}
In the last step, we used invariance of the action $S\{\phi+u,B+v\}$ under the gauge transformation of the variables $\phi+u$ and $B+v$,  and approximated $\Sigma^\mu_a[\phi+u,B+v],J^\mu_\alpha[\phi+u,B+v]$ by $\Sigma^\mu_a[\phi,B],J^\mu_\alpha[\phi,B]$. Comparison of the coefficients at $u^a_\mu$ and $v^\alpha_\mu$ finally leads to the transformation rules
\begin{equation}
\delta\Sigma^\mu_a=-f^b_{a\alpha}\Sigma^\mu_b\epsilon^\alpha,\qquad\delta J^\mu_\alpha=-f^\gamma_{\alpha\beta}J^\mu_\gamma\epsilon^\beta.
\label{currenttransfo}
\end{equation}
These demonstrate that despite the inhomogeneous transformation of the gauge field, both the scalar and the vector current are covariant under the gauge transformations.


\section{Gauge-covariant local functions}
\label{app:theoremproof}

The sole aim of this appendix is to prove in detail theorem~\ref{thm:covariantobjects}. For most of what follows, we can afford the luxury of dropping group indices. Thus, the local gauge-covariant function under consideration can be written symbolically as
\begin{equation}
\psi[\phi,A]=\psi(\phi,\de\phi,\de\de\phi,\dotsc;A,\de A,\de\de A,\dots).
\end{equation}
Although we do not insist that the fields only appear with at most one derivative attached to them (we are talking about a derivative expansion of an EFT after all), we do assume that the degree of the derivatives appearing in $\psi[\phi,A]$ is bounded from above. Take now a (possibly higher) derivative of $\phi$ and rewrite it in terms of the covariant derivatives,
\begin{equation}
\de^n\phi=\de^{n-1}(D\phi+\imag A\phi)=\de^{n-1}D\phi+\imag A\de^{n-1}\phi+\dotsb,
\end{equation}
where the ellipsis denotes terms with less than $n-1$ ordinary derivatives acting on $\phi$. Iterating this manipulation, the function $\psi[\phi,A]$ can be cast solely in terms of $\phi$ and its covariant derivatives, and of $A$ and its ordinary derivatives. This is merely a change of variables; so far we have by no means used the assumed gauge covariance. The advantage of writing $\psi[\phi,A]$ in terms of the new variables,
\begin{equation}
\psi[\phi,A]=\psi(\phi,D\phi,DD\phi,\dotsc;A,\de A,\de\de A,\dotsc),
\end{equation}
of course is that the covariant derivatives of $\phi$ do not contribute terms with derivatives of the transformation parameter $\epsilon$ to the gauge variation $\delta\psi$.

We now proceed by induction and show that the derivatives of the gauge field can all be combined into the field-strength tensor $F$ and its covariant derivatives. Start with the highest-degree derivative acting on $A$, say $\de^n A$. The gauge variation of $\psi[\phi,A]$ gets a sole contribution with $n+1$ derivatives on $\epsilon$,
\begin{equation}
\delta\psi[\phi,A]=\frac{\de\psi}{\de(\de_{\mu_1}\dotsb\de_{\mu_n}A_{\mu_{n+1}})}\de_{\mu_1}\dotsb\de_{\mu_{n+1}}\epsilon+\text{terms with less derivatives on $\epsilon$}.
\label{auxthmproof}
\end{equation}
Next introduce two tensors with a partial (anti)symmetry under the exchange of the indices $\mu_1,\dotsc,\mu_{n+1}$,
\begin{equation}
\begin{split}
\mathcal S_{\mu_1\dotsb\mu_{n+1}}&=\sum_{k=1}^{n+1}\de_{\mu_1}\dotsb\hat\de_{\mu_k}\dotsb\de_{\mu_{n+1}}A_{\mu_k},\\
\mathcal A_{\mu_1\dotsb\mu_{n+1}}&=n\de_{\mu_1}\dotsb\de_{\mu_n}A_{\mu_{n+1}}-\sum_{k=1}^n\de_{\mu_1}\dotsb\hat\de_{\mu_k}\dotsb\de_{\mu_{n+1}}A_{\mu_k},
\end{split}
\end{equation}
where the hat indicates an omitted factor in the product. The $n$-th derivative of $A$ can be expressed in terms of these tensors, $\de_{\mu_1}\dotsb\de_{\mu_n}A_{\mu_{n+1}}=(\mathcal S_{\mu_1\dotsb\mu_{n+1}}+\mathcal A_{\mu_1\dotsb\mu_{n+1}})/(n+1)$. In terms of the Abelian part of the field-strength tensor, $f_{\mu\nu}=\de_\mu A_\nu-\de_\nu A_\mu$, we also have
\begin{equation}
\mathcal A_{\mu_1\dotsb\mu_{n+1}}=\sum_{k=1}^n\de_{\mu_1}\dotsb\hat\de_{\mu_k}\dotsb\de_{\mu_n}f_{\mu_k\mu_{n+1}}.
\end{equation}
Altogether, $\de_{\mu_1}\dotsb\de_{\mu_n}A_{\mu_{n+1}}$ can be traded for a combination of $\mathcal S_{\mu_1\dotsb\mu_{n+1}}$ and terms of the type $\de_{\mu_1}\dotsb\de_{\mu_{n-1}}f_{\mu_n\mu_{n+1}}$. The latter are antisymmetric in a pair of indices and thus do not contribute to $\delta\psi$ a term with $n+1$ derivatives on $\epsilon$. From eq.~\eqref{auxthmproof} and the assumed gauge covariance of $\psi[\phi,A]$ we then infer that $\de\psi/\de\mathcal S_{\mu_1\dotsb\mu_{n+1}}=0$. Finally, re-express $f_{\mu\nu}$ in terms of $F_{\mu\nu}$ and a product of $A$'s so that the gauge-covariant function acquires the form
\begin{equation}
\psi[\phi,A]=\psi(\phi,D\phi,DD\phi,\dotsc;A,\de A,\dotsc,\de^{n-1}A;\de^{n-1}F).
\end{equation}
This argument can now be iterated. At each step, there are at most $k$ derivatives acting on each $A$ and $k$ ordinary derivatives acting on $F$ and its covariant derivatives. The latter can be reduced by expressing $\de^k F$ in terms of $\de^{k-1}DF$ and $\de^{k-1}AF$. The absence of explicit dependence on $\de^kA$ is then proved following the same steps as above. Eventually, we reach the point at which the gauge-covariant function can be written as
\begin{equation}
\psi[\phi,A]=\psi(\phi,D\phi,DD\phi,\dotsc;A;F,DF,\dotsc,D^{n-1}F).
\end{equation}
Now absence of derivatives of $\epsilon$ in $\delta\psi$ simply requires that $\de\psi/\de A=0$. This completes the proof that $\psi[\phi,A]$ can be expressed solely in terms of $\phi$, $F$ and their covariant derivatives. The question of possible gauge invariance of $\psi[\phi,A]$ thus boils down to the consideration of global symmetry transformations alone.


\section{Absence of Chern-Simons terms at order four}
\label{app:CSfour}

In this appendix, we sketch the proof that there are no nontrivial CS terms of order four in derivatives. First, the most general gauge-covariant current of order three takes the form
\begin{equation}
J^\mu_\alpha=c^{\mu\nu\kappa\lambda}_{\alpha\beta}D_\nu G^\beta_{\kappa\lambda},
\end{equation}
where $c^{\mu\nu\kappa\lambda}_{\alpha\beta}$ is without loss of generality antisymmetric in $\kappa,\lambda$. The current conservation condition~\eqref{currentconservation} takes the form
\begin{equation}
0=c^{\mu\nu\kappa\lambda}_{\alpha\beta}D_\mu D_\nu G^\beta_{\kappa\lambda}=2c^{\mu\nu\kappa\lambda}_{\alpha\beta}\de_\mu\de_\nu\de_\kappa B^\beta_\lambda+\dotsb,
\label{munualpha}
\end{equation}
where the ellipsis denotes terms with less than three derivatives acting on $B^\alpha_\mu$. Let us for the moment set $T^{\mu\nu\kappa}=c^{\mu\nu\kappa\lambda}_{\alpha\beta}$. We can always decompose this tensor into components with partial (anti)symmetry, $T^{\mu\nu\kappa}=S^{\mu\nu\kappa}+A^{\mu\nu\kappa}+U^{\mu\nu\kappa}+V^{\mu\nu\kappa}$, where
\begin{align}
\notag
S^{\mu\nu\kappa}&=\tfrac16(T^{\mu\nu\kappa}+T^{\mu\kappa\nu}+T^{\nu\kappa\mu}+T^{\nu\mu\kappa}+T^{\kappa\mu\nu}+T^{\kappa\nu\mu}),\qquad\text{symmetric in $\mu,\nu,\kappa$,}\\
\notag
A^{\mu\nu\kappa}&=\tfrac16(T^{\mu\nu\kappa}-T^{\mu\kappa\nu}+T^{\nu\kappa\mu}-T^{\nu\mu\kappa}+T^{\kappa\mu\nu}-T^{\kappa\nu\mu}),\qquad\text{antisymmetric in $\mu,\nu,\kappa$,}\\
U^{\mu\nu\kappa}&=\tfrac13(T^{\mu\nu\kappa}+T^{\nu\mu\kappa}-T^{\mu\kappa\nu}-T^{\kappa\mu\nu}),\qquad\text{antisymmetric in $\nu,\kappa$,}\\
\notag
V^{\mu\nu\kappa}&=\tfrac13(T^{\mu\nu\kappa}+T^{\mu\kappa\nu}-T^{\nu\mu\kappa}-T^{\nu\kappa\mu}),\qquad\text{antisymmetric in $\mu,\nu$.}
\end{align}
The condition~\eqref{munualpha} ensures that $S^{\mu\nu\kappa}=0$. Moreover, both $A^{\mu\nu\kappa}$ and $U^{\mu\nu\kappa}$ are antisymmetric in $\nu,\kappa$, giving rise to $c^{\mu\nu\kappa\lambda}_{\alpha\beta}$ which is fully antisymmetric in $\nu,\kappa,\lambda$ and thus drops from the current thanks to the Bianchi identity. The whole tensor $T^{\mu\nu\kappa}$ can therefore be replaced with its part $V^{\mu\nu\kappa}$, which finally implies that $c^{\mu\nu\kappa\lambda}_{\alpha\beta}$ must be antisymmetric in $\mu,\nu$. As a consequence, the conservation condition~\eqref{munualpha} becomes simply
\begin{equation}
c^{\mu\nu\kappa\lambda}_{\alpha\beta}f^\beta_{\gamma\delta}G^\gamma_{\mu\nu}G^\delta_{\kappa\lambda}=0,
\end{equation}
and is satisfied if and only if $c^{\mu\nu\kappa\lambda}_{\alpha\beta}f^\beta_{\gamma\delta}$ is symmetric under the exchange of $\mu,\nu$ and $\kappa,\lambda$. Moreover, $c^{\mu\nu\kappa\lambda}_{\alpha\beta}$ must be an invariant tensor of (the adjoint representation of) $H$, which means among others that $c^{\mu\nu\kappa\lambda}_{\alpha\beta}f^\beta_{\gamma\delta}$ is antisymmetric in the three indices $\alpha,\gamma,\delta$ (see appendix~\ref{app:tensor} for details). It is now a matter of a straightforward, if somewhat tedious, calculation to show that the corresponding Lagrangian, defined by eq.~\eqref{CSmaster}, is up to a total derivative equal to $\frac14c^{\kappa\lambda\mu\nu}_{\alpha\beta}G^\alpha_{\kappa\lambda}G^\beta_{\mu\nu}$. Hence the Lagrangian is necessarily gauge-invariant, and there are no nontrivial CS terms at this order.


\bibliographystyle{JHEP}
\bibliography{references}


\end{document}